\documentclass[12pt]{iopart}

\usepackage{graphicx}
\usepackage{subcaption}
\usepackage{bm}
\usepackage{esint}
\usepackage{xcolor}

\newcommand{\bra}[1]{\ensuremath{\left\langle#1\right\vert}} 
\newcommand{\ket}[1]{\ensuremath{\left\vert#1\right\rangle}} 
\newcommand{\melem}[3]{\ensuremath{\left\langle#1\middle|#2\middle|#3\right\rangle}} 
\newcommand{\braket}[2]{\ensuremath{\left\langle#1\middle|#2\right\rangle}} 
\newcommand{\operator}[1]{\ensuremath{\widehat{#1}}} 
\renewcommand{\vector}[1]{\ensuremath{\bm{#1}}} 
\newcommand{\of}[1]{\ensuremath{\left(#1\right)}} 
\newcommand{\abs}[1]{\ensuremath{\left\vert#1\right\vert}} 
\newcommand{\im}{\ensuremath{\mathrm{i}}} 

\begin{document}

\title[Adiabatic passage in tripod systems]{Study of the adiabatic passage in tripod atomic systems in terms of the Riemannian geometry of the Bloch sphere.}

\author{Arturs~Cinins}
\address{Institute of Atomic Physics and Spectroscopy, University of Latvia, Jelgavas str. 3, LV-1004 Riga, Latvia}
\ead{arturs.cinins@lu.lv}

\author{Martins~Bruvelis}
\address{King Abdullah University of Science and Technology (KAUST), Computer, Electrical and Mathematical Sciences and Engineering Division (CEMSE), Thuwal 23955-6900, Saudi Arabia}

\author{Nikolai~N.~Bezuglov}
\address{Saint Petersburg State University, 7/9 Universitetskaya nab., St. Petersburg 199034, Russia}
\address{Rzhanov Institute of Semiconductor Physics SB RAS, Novosibirsk 630090, Russia}

\vspace{10pt}
\begin{indented}
\item[]May 2022
\end{indented}

\begin{abstract}

We present an analysis of the stimulated Raman adiabatic passage processes based on the methods of differential geometry.
The present work was inspired by an excellent article by Bruce W.~Shore et~al. (R.~G.~Unanyan, B.~W.~Shore, and K.~Bergmann Phys.~Rev.~A~\textbf{59}, 2910 (1999)).
We demonstrate how a purely geometric interpretation of the adiabatic passage in quantum tripod systems as a Riemannian parallel transport of the dark state vector along the Bloch sphere allows describing the evolution of the system for a given sequence of Stokes, pump and control laser excitation pulses.
In combination with the Dykhne-Davis-Pechukas adiabaticity criterion and the minimax principle for circles on a sphere, this approach allows obtaining the analytical form of the optimal laser pulse sequences for a high fidelity tripod fractional STIRAP.
In contrast to the conventional STIRAP in $\Lambda$-systems, the Gaussian approximations of the optimal laser pulse sequences allow reaching the infidelity of $10^{-7}$ for the adiabaticity parameter of $300$ without noticeable oscillatory or other detrimental effects on population transfer accuracy.
\end{abstract}

%
%
%
%
%

\section{Introduction}
\label{sect:Introduction}

Stimulated Raman adiabatic passage (STIRAP) is a robust method for selective population transfer between quantum states with many applications in modern physics, chemistry, and information processing \cite{Bergmann_2019,Shore2017}.
STIRAP processes in tripod systems \cite{Unanyan_1999, Vitanov2017} with  their generalization on N-pod quantum systems \cite{Vitanov2017} is of particular interest for quantum information because of the two (or N-1) orthogonal dark states that can form a qubit (or so called qudit) \cite{Vitanov2013}.

In atomic systems containing $n$ degenerate adiabatic states $\ket{D_i}$ (d-states) with constant energy $\varepsilon_d$ independent of the slowly varying parameters $\Re_ n$ of the system, the adiabatic passage  has a number of specific features.
These features are mainly due to the presence of irremovable transitions in the subspace $\Lambda _n$  of d-states caused by the operator of nonadiabatic coupling.
The efficiency of the corresponding non-adiabatic mixing of d-states does not depend on the temporal scales of time-dependent system parameters and, as noted in the physical literature, is determined by the geometry of the parameter space, more precisely, the topology of a closed curve $\Re_ n(t)$  formed by the parameters upon a complete adiabatic cycle \cite{Berry_1984}.
At the same time, it was established that the temporal dynamics of d-states are reduced to a group  $\operator{U}_{\Re }$ of unitary transformations in the subspace $\Lambda _n$ and that gauge fields are the appropriate tool for describing such transformations \cite{Wilczek1984}.
To the best of our knowledge, the first study of gauge structure for tripod STIRAP was made in a paper by Bruce W.~Shore et~al. \cite{Unanyan_1999}, where the operators $\operator{U}_{\Re }$ were shown to form the orthogonal rotation group SO($2$) of the two-dimensional dark states subspace $\Lambda_2$ with final evolution given by the value of the so-called geometric phase \cite{Berry_1984}.

\begin{figure}[ht]
    \centering
    \begin{subfigure}[b]{0.3\textwidth}
        \centering
        \includegraphics[width=\textwidth]{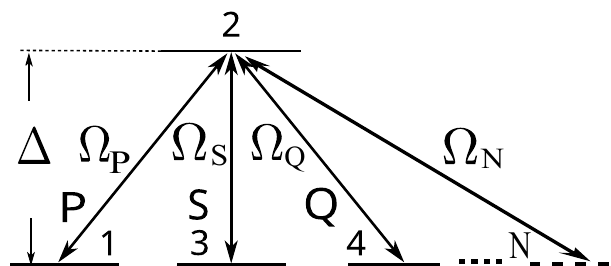}
        \caption{}
        \label{fig:tripod_bare}
    \end{subfigure}
    \hspace{10pt}
    \begin{subfigure}[b]{0.3\textwidth}
        \centering
        \includegraphics[width=\textwidth]{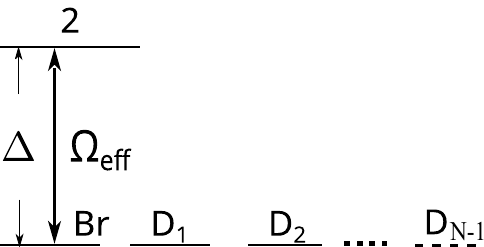}
        \caption{}
        \label{fig:tripod_dark_bright}
    \end{subfigure}
    \caption{
        Energy levels in  tripod (N-pod) systems under the rotating wave approximation with  
        (a) lasers $P, S, Q$ $(,\ldots)$ coupling the ground sublevels $1,3,4$ $(,\ldots,N)$ to the excited state $2$.
        The parameter $\Delta$ corresponds to the single-photon detuning.
        (b) Reduction of the linkage diagram (a) to a single coupled bright state $Br$ and a set of decoupled dark states $D_{1,2(,\ldots,N-1)}$.
        Dashed lines represent the additional sublevels $\ket{j}, j>4$ for N-pod systems with $N>3$.
    }
    \label{fig:tripod}
\end{figure}

In this paper, we consider a somewhat different, more geometric approach to solving the problems of adiabatic passage based on the methods of differential and Riemannian geometries \cite{Kreyszig1991, Arnold_1978}.
We proceeded from the remark made in the book \cite{Fadeev_1980} on interpreting gauge fields as the cause of the curvature of the so-called charge space, which for tripod systems is a kind of analog of d-subspace $\Lambda_{2}$.
The corresponding linkage diagram is shown in Fig.~\ref{fig:tripod_bare} along with the excitation scheme (Fig.~\ref{fig:tripod_dark_bright}) reducing to two dark and one bright state (see details in Section~\ref{sect:Bright21}).
In Section~\ref{sect:Geometrical22}, we will demonstrate that the normalized bright state $\ket{Br}$ can be associated with a unit vector $\vector{e}_R=\vector{R}/|\vector{R}|$, where the vector  $\vector{R}\sim (\Omega _P, \Omega _S, \Omega _Q)$ (Rabi vector) belongs to the three-dimensional parameter space $\Re_3$ of laser Rabi frequencies $\Omega _i$ ($i=S,P,Q$).
Simultaneously, the unit vectors associated with  dark states $\ket{D}$ lie in the two-dimensional plane $\Lambda_{R}$ tangent to the unit sphere at point $\vector{e}_R$.
Temporal evolution of  $\Omega _i (t)$ determines a path $\vector{e}_R(t)$ (Rabi path) on the surface of a unit sphere (analogous to the Bloch sphere), while temporal evolution of the dark state-vector due to nonadiabatic coupling is a direct consequence of Riemannian parallel transport \cite{Fadeev_1980} of the ”dark” tangent planes $\Lambda _R$ along the Rabi path.
Based on a purely geometric approach, in Section~\ref{sect:Geometrical3} analytical expressions are obtained for optimal sequences of laser pulses that satisfy the Dykhne-Davis-Pechukas adiabaticity criterion $|\vector R(t)|=const$ \cite{Unanyan_1999, Dykhne1960, DavisPechukas1976} and implement a fractional STRAP where only a controlled fraction of population transfer occurs.
Section~\ref{sect:Examples} presents a series of numerical simulations of the tripod quantum dynamics in the case of the most characteristic STIRAP processes with optimal lasers pulse trains.
Importantly, their temporal profiles allow for Gaussian approximations that are convenient for experimental implementation, while providing good population transfer accuracy with infidelity reaching $10^{-7}$ for the adiabaticity parameter of $300$.

Since our work is intended for a physical audience and does not imply knowledge of the fundamentals of differential geometry, we have provided the description and elementary proofs of the necessary provisions to the Appendices.
In \ref{appendix:NonadiabaticA}, the problem of the dark state evolution is considered within two different frameworks: non-adiabatic coupling due to quantum-mechanical effects and the Riemannian procedure of parallel transport of tangent planes.
It is shown that a quantitative analysis of the adiabatic passage in both approaches results in identical analytical expression (\ref{eq:StateVectorEvolutionFromQuantumMechanics}) for the local rotations of  two-dimensional dark subspaces $\Lambda_2$.
\ref{appendix:BerryPhase} is devoted to finding a convenient expression (\ref{eq:ContourIntegeralEq9}) for the geometric factor $\beta$ that determines the fraction of the initial state population transferred in the STIRAP process in terms of a contour integral over a closed-loop  in the parameter space $\Re_3$.
An interesting feature of the resulting expression (\ref{eq:ContourIntegeralEq9}) for $\beta$ is related to the Dirac vector potential (\ref{eq:ContourIntegeralEq9}) included in it, generated by a unit magnetic charge \cite{Dirac1931}.  The last \ref{appendix:InfidelityC} contains the mathematical details on the derivation of the excited state probability amplitude (\ref{eq:AmplitudeC_2}) along with formula (\ref{eq:InfidelityE40}) for the infidelity parameter of fractional STRAP.

Noteworthy, in contrast to most works that study the problems of tripod systems and specify an artificial basis for dark states, a natural dynamic set of dark basis states emerges from our approach, associated with geometric properties of curves in the parameter space of the laser Rabi frequencies.
The geometric approach for determining the evolution of degenerate dark states during adiabatic passage can also be generalized to N-pod atomic systems.
Therefore, the wording of some provisions is given for the case $N>3$ where appropriate.

\section{Notation, assumptions and remarks}
\label{sect:Notation}
Preparation of quantum objects into a proper initial state with subsequent transfer to another predefined state is among the fundamental tasks of quantum optics and informatics.
One approach to solve such problems is to directly apply fractional STIRAP to the tripod (or N-pod) systems with the energy levels diagram depicted in Fig.~\ref{fig:tripod}.
Specifically, we are concerned with producing a coherent superposition of states \ket{1} and \ket{3} when the system is initially (at $t=t_0$) in state \ket{1}:
\begin{eqnarray}
    \label{eq:E1}
    \psi (t=t_0)=\ket{1} \rightarrow  \psi (t=t_f)= \cos{\beta}\ket{1}+\cos{\beta}e^{\im\gamma}\ket{3}.
\end{eqnarray}
We are dealing with one excited quantum state $\ket{2}$ and $N$ stable components $\ket{i}$ ($i=1,3,...N$) of the ground state, which are subject to interaction with $N$ laser fields.
The pump (P) and Stokes (S) lasers drive the level population transfer, while the other $N-2$ control lasers play an auxiliary role.

For a given sequence of laser pulses, an exact analysis of the transformation~(\ref{fig:tripod}), parameterized by the mixing angle $\beta$ and the relative phase $\gamma$, requires solving the Schrödinger equation, which under the rotating wave approximation (RWA) reads:
\begin{eqnarray}
    \label{eq:E2}
  \im\frac{d}{dt} \psi =\frac{1}{\hbar}\operator{H}\psi; \quad \operator{H}=\varepsilon _{2}\ket{2}\bra{2} + \operator{V}; \\
  \label{eq:E3}
 \operator{V}= \frac{\hbar}{2}\sum_{j=1,3,..}^{N}\Omega _{j}(t) \ket{2}\bra{j}+h.c.
\end{eqnarray}
The first term in the operator $\operator{H}$ (\ref{eq:E2}) determines the energy structure of the N-pod system: the energy $\varepsilon _{g}$ of all degenerate bare ground sublevels $\ket{j}$ is taken equal to zero, while the energy $\varepsilon _{2}$  of the upper bare state $\ket{2}$ corresponds to the single-photon detuning $\Delta$ of lasers: $\varepsilon _{2}=\hbar \Delta$.
The operator $\operator{V}(t)$ (\ref{eq:E3}) describes the atomic levels coupling with the laser fields via their slow varying Rabi frequencies $\Omega _{j}=(2/\hbar)\bra{2}\operator{V}\ket{j}$.
According to (\ref{eq:E1}), the initial condition for the state vector $\psi$ implies $\psi (t=t_{0}) =\ket{1}$.

The key point for our approach is the ability to assume, without a loss of generality, that
(i)  Rabi  frequencies  $\Omega _{j}$  of all lasers  are real and
(ii) the relative phase $\beta$ (\ref{eq:E1}) is equal to zero (for the relevant rationale see \cite{Unanyan1998} when discussing the formulas (4-7) given there).

\subsection{Bright and dark states in N-pod systems}
\label{sect:Bright21}
The implementation of passage (\ref{eq:E1}) without uncontrolled phase losses assumes the absence of mixing, generated by the field operator $\operator{V}$ (\ref{eq:E3}),  between the state-vector $\psi$ and the unstable upper level $2$.
To avoid possible dephasing processes, one needs to support the embedding of the vector $\psi(t)$ in the subspaces  $\Lambda _D (t)$ of dark states, which become time-dependent when the Rabi frequencies alter. The criterion for the vector
\begin{eqnarray}
    \label{eq:E4}
 \ket{D}=\sum_{j\neq 2}^{}C_{j}^{(D)}\ket{j}
\end{eqnarray}
to belong to the category of dark states reduces to zeroing the matrix element $\bra{2}\operator{V}\ket{D}$ that means \cite{Kirova2017}:
\begin{eqnarray}
    \label{eq:E5}
 \bra{2}\operator{V}\ket{D}= \frac{\hbar}{2} \sum_{j\neq 2}^{}\Omega _{j}C_{j}^{(D)}=0.
\end{eqnarray}
Equations (\ref{eq:E4}), (\ref{eq:E5}) may be treated as an orthogonal condition
\begin{eqnarray}
    \label{eq:E6}
  \left<Br|D \right>=0; \quad \ket{Br}= \sum_{j\neq 2}^{}\Omega _{j}\ket{j}/\sqrt{\sum_{j\neq 2}\Omega _{j}^2}
\end{eqnarray}
between any dark state (\ref{eq:E4})  and the newly introduced unit wave vector $\ket{Br}$ (\ref{eq:E6}).
Straightforward calculation yields
\begin{eqnarray}
    \label{eq:E7}
 \frac{1}{\hbar}\bra{2}\operator{V}\ket{Br}=\frac{1}{2}\sqrt{\sum_{j\neq 2}\Omega _{j}^2}\equiv \frac{1}{2}\Omega _{eff},
\end{eqnarray}
i.e. the unit vector $\ket{Br}$ appears, in contrast to all decoupled dark $\ket{D}$ substates, to be strongly coupled to the excited state $2$, the linkage constant $\Omega _{eff}$ having played the role of effective Rabi frequency.
For this reason, the state $\ket{Br}$ can be termed "bright".

Equation (6) implies that the subspace $\Lambda _D$, composed from the dark states is orthogonal to the
one-dimensional subspace $\Lambda _{Br}$, containing the single bright state $\ket{Br}$. The dimension of $\Lambda _D$, thus, is $N-1$, i.e., one can choose $N-1$ mutually orthogonal dark states $\ket{D_k}$. The
corresponding linkage diagram for the the field operator  $\operator{V}$ (3) in the basis $ \ket{Br}, \ket{D_k}$ is depicted in Fig.~\ref{fig:tripod}(b). Noteworthy, upon altering Rabi frequencies, both subspaces $\Lambda _{Br}(t),\Lambda _D(t)$ become time-dependent.

The N-pod operator $\operator{H}$ (2) acts independently in the subspace $\Lambda _D$  of dark states and subspace $\Lambda _{\pm }$ of  two coupled  $\ket{Br},\ket{2}$ vectors. Diagonalization
of $\operator{H}$ in $\Lambda _{\pm }$ results in the formation of two
adiabatic (dressed) states $\ket{\pm }$  as superpositions of vectors $\ket{Br}$ and $ \ket{2}$ \cite{Unanyan1998,Kirova2017} with   repulsive adiabatic energies
\begin{eqnarray}
    \label{eq:E8}
\varepsilon _{\pm }(t) =\frac{\hbar}{2}\Delta \pm\frac{\hbar}{2}\sqrt{\Delta ^2+\Omega _{eff}(t)^2}
\end{eqnarray}
that alter  as the laser pulses pass. At the same time, the dark states $\ket{D}$ obey the relation $\operator{H}\ket{D}=0$, i.e. all $\ket{D}$  are degenerate adiabatic states with zero energy $\varepsilon _{D}\equiv 0$, regardless of laser coupling strengths.

Although the four-level system ($N=3$), which we will focus on below, will be sufficient to perform an operation (\ref{eq:E1}), increasing the number of degrees of freedom makes N-pod systems more flexible and allows more complex quantum transformations \cite{Vitanov2013}.
Importantly, the configurations of dark states, depicted in Fig.~\ref{fig:tripod_dark_bright} are of the same type so that the basic ideas of the geometric approach as applied to tripod STIRAP also work in the general situation with $N>3$.

\subsection{Geometrical counterparts of Bright and Dark states}
\label{sect:Geometrical22}

Since all the coefficients in relations (\ref{eq:E4}-\ref{eq:E7}) are real, we can associate the set of Rabi frequencies $\Omega _{j}$ that define the bright state wave function $\ket{Br}$ (\ref{eq:E6}) as components of the Rabi
 vector $\vector{R}\sim (\Omega _P, \Omega _S, \Omega _Q,...,\Omega _{N})$ in some Euclidean parameter  space $\Re_ N$.
In this case, if we assume that the probability amplitudes $C_{j}$ for dark quantum states $\ket{D}$ (\ref{eq:E4}) also form the Euclidean vector  $\vector{D}\sim (C _P, C _S, C _Q,...,C _{N})$ , then the quantum scalar product turns into the usual scalar product of Euclidean vectors: $\left<Br|D \right>=(\vector{R}\vector{D})$.

In what follows, as applied to tripod systems, we will identify the unit vectors $\vector{e}_{Q},\vector{e}_{S},\vector{e}_{P}$ of the Euclidean three-dimensional space $\Re _3$ with the basis bare states  $\ket{4},\ket{3},\ket{1}$ respectively (Fig.~\ref{fig:ParallelTransport}). In other words, the following correspondence occurs:
\begin{eqnarray}
    \label{eq:E9}
\vector{\Psi}=C_4\ket{4}+C_3\ket{3}+C_1\ket{1} \Leftrightarrow \vector{A}=C_4\vector{e}_{Q}+C_3\vector{e}_{S}+C_1\vector{e}_{P}
\end{eqnarray}
in the case of dark (tangent) vectors and
\begin{eqnarray}
    \label{eq:E10}
\vector{\psi}=\Omega _Q\ket{4}+\Omega _S\ket{3}+\Omega _P\ket{1} \Leftrightarrow \vector{R}=\Omega _Q\vector{e}_{Q}+\Omega _S\vector{e}_{S}+\Omega _P\vector{e}_{P}
\end{eqnarray}
to denote vectors associated with Rabi frequencies of lasers (Rabi vectors $\vector{R}$) and bright states ($\vector{e}_{R}=\vector{R}/\Omega _{eff}$).

\subsection{Remarks on non-adiabatic transitions in terms of quantum and geometrical approaches}
\label{sect:RemaksTransitions23}

With an adiabatic change in the Rabi frequencies, the Rabi vector $\vector{R}(t)$ (\ref{eq:E10})  moves along some curve $\Im _R$ in the specified parameter space $\Re_ N$. At the same time, the bright vector $\vector{e}_R(t) =\vector{R}/|\vector{R}|$ (\ref{eq:E6}) traverses the Rabi path $\Im _1=\operator{\Xi} _{R\to 1}\Im _R$,  which is the radial projection of the curve $\Im _R$ on the unit generalized Bloch sphere of $(N-1)$-dimension embedded in parameter space $\Re_ N$ (Fig.~\ref{fig:ParallelTransportXYplane}).
Since the subspace of dark states is orthogonal to the bright state, at time $t$ all dark states vectors must lie in a plane $\Lambda _{D}(t)$ tangent to the unit sphere at $\vector{e}_R(t)$.

Everywhere below, we will assume the feasibility of the adiabaticity criterion, which reduces to requiring large values of the so-called adiabaticity parameter or pulses area \cite{Unanyan_1999,Vitanov2017}:
\begin{eqnarray}
\label{eq:E11}
\wp = \frac{1}{2}\int_{-\infty }^{\infty}\Omega _{eff}(t)dt = \frac{1}{2}\int_{-\infty }^{\infty}\sqrt{\sum_{j\neq 2}\Omega _{j}^2} dt    \gg 1.
\end{eqnarray}
Since the blocking of unwanted non-adiabatic transitions between states
$\ket{\pm}$ and $\ket{D}$ is controlled by increasing their splitting energy $\varepsilon _{\pm}$~(\ref{eq:E8}), which includes $\Omega _{eff}/2$, we keep the one-half factor in the definition (\ref{eq:E11}).
Perfect adiabatic passage prevents population flow between dark and bright states and allows interpreting degenerate dark states as an independent closed system of levels with a source of quantum transitions between them in the form of a nonadiabatic coupling operator \cite{Wilczek1984}.

As was mentioned in the introductory Section~\ref{sect:Introduction},  the authors of a number of papers \cite{Unanyan_1999,Berry_1984,Wilczek1984} propose to analyze the evolution of dark states in terms of gauge fields theory. On the other hand, the book \cite{Fadeev_1980} indicates a deep connection between gauge fields and the curvature of the configuration space of the matter carriers of the fields themselves. As applied to the adiabatic passage, this means that when a bright vector traverses a small segment of the Rabi curve $\Im _1$ on a Bloch sphere, the  Riemannian  parallel transport of tangent (i.e., dark) vectors along the segment is equivalent to the action of a nonadiabatic coupling on them.

The justification for this equivalence is given in \ref{appendix:NonadiabaticA} for tripod STIRAP, where the tangent planes $\Lambda _D$  to the Bloch sphere are two-dimensional, and the evolution of dark vectors as the segment $d\vector{e}_R$ passes is reduced to their rotation by an angle $d \beta _s $ (\ref{eq:AngleA12}) in the parameter space $\Re_ 3$ (Fig.~\ref{fig:ParallelTransportXYplane}).
After the end of the laser pulses action, the $\Im _R$ curve and its Rabi image $\Im _1$ on the Bloch sphere return to the initial points, while the dark vector rotates through a certain angle $\beta$ (\ref{eq:E1}), called the geometric phase or geometric factor \cite{Unanyan_1999,Vitanov2017}. The famous Riemann theorem expresses the geometric factor with the area of a surface element (solid angle) cut by  closed curve $\Im _1$ on the Bloch sphere \cite{Kreyszig1991,Arnold_1978}. \ref{appendix:BerryPhase} provides a more convenient expression~(\ref{eq:ContourIntegeralEq8}) for the angle $\beta$ through the one-dimensional contour integral
\begin{eqnarray}
\label{eq:E12}
    \beta &= \oint_{\Im }\frac{\Omega_S d\Omega_P - \Omega_P d\Omega_S }{\Omega_{eff}(\Omega_{eff}+\Omega_Q)}.
\end{eqnarray}
over the cycle $\Im $, which can be chosen as the close path $\Im _R$ associated with  the Rabi vector $\vector{R}(t)$ in the parameter space, or as a Rabi curve $\Im _1=\operator{\Xi } _ {R\to 1}\Im _R$ on the Bloch sphere.

\subsection{Remarks on optimization of laser pulses}
\label{sect:RemaksOptimization24}
The accuracy of performing operations~(\ref{eq:E1}) is significantly affected by two factors. The first of them is directly related to the time scale of laser pulses and regulates the loss of STIRAP efficiency due to the uncontrolled nonadiabatic coupling of dark states with bright ones. Minimization of this loss is achieved by matching the parameters of laser pulses using the Dykhne-Davis-Pechukas (DDP) adiabaticity condition: the value of the effective Rabi frequency~(\ref{eq:E7})  must remain constant. Indeed, the energies $\varepsilon _{\pm }$ (8) and $\varepsilon _{D}$ of all adiabatic states do not depend on time and, therefore, with analytic continuation of the time variable $t$ to the complex plane, it is impossible to find any intersection Landau-Zener points of the functions $\varepsilon _{\pm }(t)$ and $\varepsilon _{D}(t)$, where nonadiabatic transitions are allowed \cite{Landau_QM1981}. In terms of geometric interpretation, any optimal DDP sequence $\Omega _j(t)$ of laser pulses is formed by a certain closed curve $\Im _1$, determined by $\vector{e}_R(t)$ on the Bloch sphere:
\begin{eqnarray}
    \label{eq:E13}
\vector{\psi}(t)= \overline{\Omega }\vector{e}_R(t); \quad \vector{e}_R(t)=\widetilde{\Omega} _Q(t)\vector{e}_{Q}+\widetilde{\Omega} _S(t)\vector{e}_{S}+\widetilde{\Omega} _P(t)\vector{e}_{P}
\end{eqnarray}
where the notation (10) is used.  The unit  vector $\vector{e}_R(t)$ containing reduced  Rabi frequencies $\widetilde{\Omega} _j$ defines the pulses shapes, while the parameter $\overline{\Omega} $ gives the common value of Rabi frequencies. In fact, by definition (\ref{eq:E7}), $\overline{\Omega}= \Omega_{eff}$. The DDP criterion guarantees a high STIRAP fidelity with a variation of the adiabaticity parameter ~(\ref{eq:E11}) $\wp\sim \overline{\Omega }T$ where $T$ is a duration of lasers pulses.

Another factor of optimization is related to choosing the optimal  shape of the loop  $\Im _1$ in the form of a circle $C_{s}$.
This choice is due to the minimax principle: among all closed curves on the Bloch sphere with a common initial (and hence final) point and having a fixed length, a circle bounds the surface with the smallest area \cite{Kreyszig1991,Arnold_1978}, i.e. with the smallest geometric factor. Therefore, in accordance with the variational principle \cite{Arnold_1978}, the fractional STIRAP (\ref{eq:E1}) acquires stability with respect to small uncontrolled fluctuations in the parameters of laser pulses (intensities, relative temporal profiles, etc.). Noteworthy, the optimal "circular" pulses can be sufficiently well approximated (see the corresponding discussion in subsection~\ref{sect:Simulations41}) by Gaussian pulses most suitable for practical applications.

Once more remark concerns the fact of the geometric factor (\ref{eq:E12}) independence  on both the time scale and the specific parametrization of the loops $\Im _1$, $\Im _R$. Usually, the method of parameterization is dictated by the geometric properties of the loop, for example, in the case of a circle $C_s$ shown in Fig.\ref{fig:ParallelTransportXYplane}, an appropriate parameter is the azimuth angle $\varphi$. If the pulses have a finite duration (Gaussian pulses, for instance), then it is convenient to take as a parameter the dimensionless value $\tau (t) =\pi \int_{-\infty }^{t} \Omega _{eff}(\widetilde{t})d\widetilde{t} /\wp $, which varies from zero to the $2\pi$ and may be treated as the reduced angle $\varphi$.  Noteworthy, an important feature of the integrand (\ref{eq:E12}): its numerator vanishes at the point $\vector{R}=\overline{\Omega} \vector{e}_Q$ where $\Omega _S=\Omega _P=0$. Since the start and endpoints of a closed Rabi curve on the Bloch sphere must match $\vector{e}_Q$ to perform fractional STIRAP (\ref{eq:E1}) (Fig.~\ref{fig:ParallelTransportSTIRAP} and the discussion below in the next section), the value of the geometric factor $\beta $ does not critically depend on the relative pulse shapes at the initial and final phases of adiabatic passage.  This fact allows one to approximate the optimal trigonometric laser pulses with realistic Gaussian profiles (\ref{eq:OmegaQgaussian})-(\ref{eq:OmegaPgaussian}).

\section{Geometrical solutions for optimal fractional STIRAP in tripod systems}
\label{sect:Geometrical3}

\begin{figure}[ht]
    \centering
    \begin{subfigure}[b]{0.47\textwidth}
        \centering
        \includegraphics[width=\textwidth]{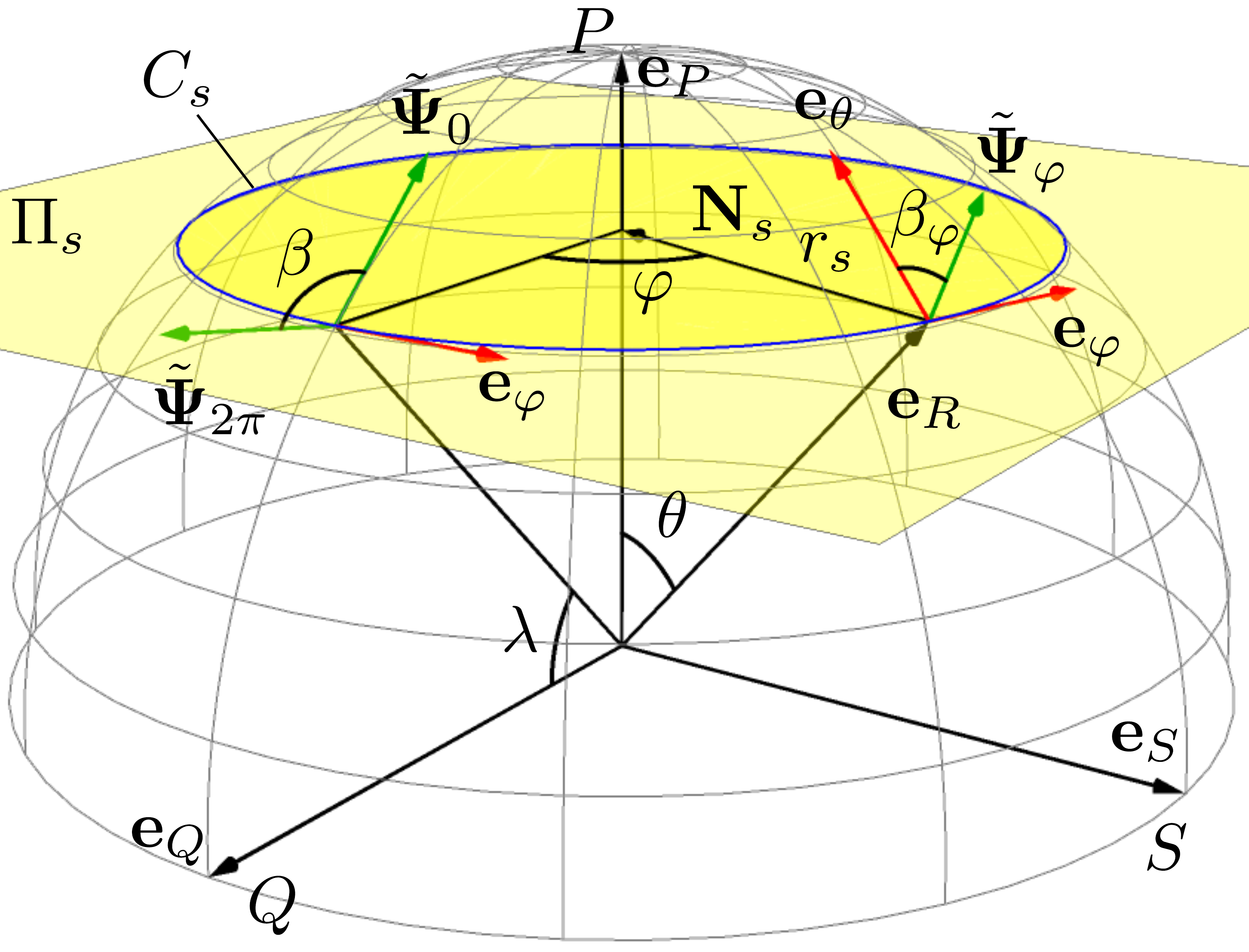}
        \caption{}
        \label{fig:ParallelTransportXYplane}
    \end{subfigure}
    \hspace{10pt}
    \begin{subfigure}[b]{0.46\textwidth}
        \centering
        \includegraphics[width=\textwidth]{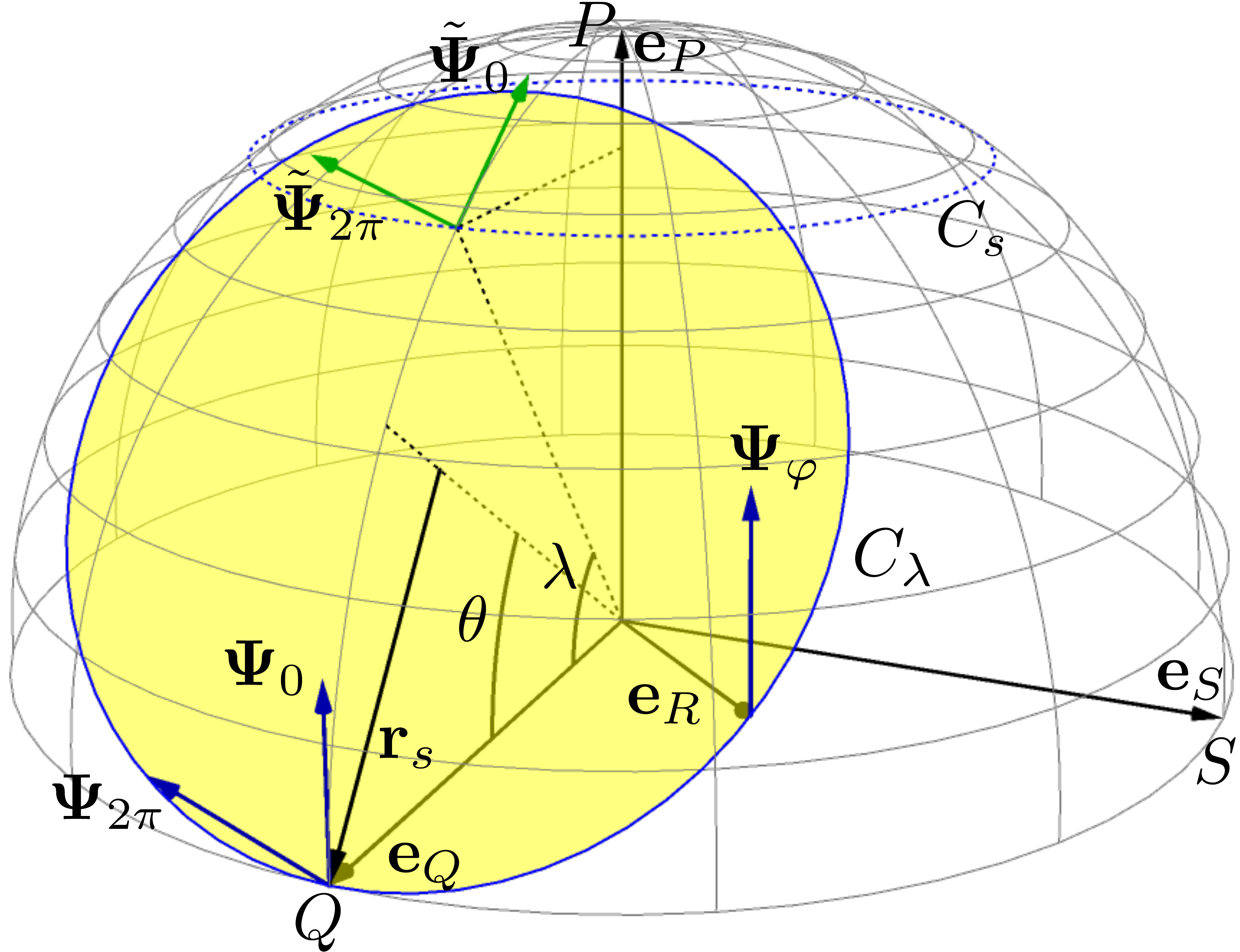}
        \caption{}
        \label{fig:ParallelTransportSTIRAP}
    \end{subfigure}
    \caption{
      Geometric interpretation of the tripod-STRAP in the form of parallel transport of tangent vectors along circular trajectories on the Bloch sphere.
        (a) The reference circle  $C_s$ lies  in a plane  parallel to the $QS$-coordinate plane.
        (b) The actual circular Rabi trajectory $C_{\lambda}$ contains the initial dark state $\Psi(\varphi=0)=\ket{1}$ along with the target state $\Psi(\varphi=2\pi)$ (\ref{eq:StateAfterFractionalSTIRAP}) in the tangent plane $\Lambda_D(\varphi=0)$ of the Bloch sphere at point $\vector{e}_Q$.
        For convenience, all normalized to unity dark (tangent) vectors are shown reduced in length.
    }
    \label{fig:ParallelTransport}
\end{figure}

We aim to implement optimal fractional tripod-STIRAP, where the initially populated state ${\ket{\Psi_0} = \ket{\Psi}(\varphi=0) = \ket{1}}$ evolves into the mixture
\begin{eqnarray}
    \label{eq:StateAfterFractionalSTIRAP}
    \ket{\Psi}(\varphi=2\pi) = \cos{\beta}\ket{1} + \sin{\beta}\ket{3},
\end{eqnarray}
with the mixing angle $\beta $ equal to the geometric phase (\ref{eq:E12}).
Assuming that perfectly adiabatic passage prevents population flow between dark and bright states, the state-vector $\ket{\Psi}(\varphi)$ must be always kept within the dark subspace.
Its geometric representation $\vector{\Psi}(\varphi)$ then always remains in the tangent space $\Lambda _D (\varphi)$ to the Bloch unit sphere at point $\vector{e}_R(\varphi)$.
After a full revolution along  optimal circle trajectory $C_{\lambda }$ (Fig.~\ref{fig:ParallelTransportSTIRAP}), the final and initial tangent planes coincide, therefore the plane $\Lambda _D (\varphi=0)$ must contain both vector representations of $\ket{1}$ and $\ket{3}$ states, associated with the unit vectors $\vector{e}_P$ and $\vector{e}_S$ accordingly (see the corresponding notations in  subsection~\ref{sect:Geometrical22}).
This is only possible if the starting point $\vector{e}_R(0)$ coincides with the unit vector $\vector{e}_Q$ corresponding  to wave vector $\ket{4}$  (Fig.~\ref{fig:ParallelTransportSTIRAP}).

\subsection{Reference optimal circles $C_{\lambda}$}
\label{sect:ReferenceTransport31}

Any circle on the Bloch sphere provides optimal laser pulses trains. The geometrical factor depends only on the circles' radius $r _s$
and, as follows from formula~(\ref{eq:E12}), it can be explicitly calculated in the case of the  circles' planes $\Pi _s$
oriented parallel to the coordinate plane $(Q,S)$ as it depicted in Fig.~\ref{fig:ParallelTransportXYplane}:
\begin{eqnarray}
    \label{eq:betaRelationTolambda}
    \beta &= 2\pi\left(1-\sin{\lambda}\right); \quad  r_s=\cos\lambda.
\end{eqnarray}

Note the following points: (i) all the lengths of the fragments in Figs.~\ref{fig:ParallelTransport} refer to the unit radius of the Bloch sphere, i.e. are dimensionless. (ii)  The angle  between master planes containing the actual $C_{\lambda}$ and the reference $C_{s}$ circles is  $\lambda$. Thus, the circle $C_{\lambda}$ can be obtained by rotating $C_s$ around  $S$-axis  through the angle $\lambda $:
\begin{eqnarray}
    \label{eq:trajectoriesRelationViaRotationGroupGenerator}
   C_{\lambda} = \exp\left(\lambda\operator{L}_S\right)C_s;\quad
    \operator{L}_j\vector{R} = \vector{e}_j\times\vector{R}
    \quad (j=Q,S,P)
\end{eqnarray}
The linear anti-symmetric
operator $\operator{L}_j$ is known in classical mechanics as the $j$-axis rotation group generator \cite{Arnold_1978}.
(iii) The rotation operator $\exp\left(\lambda\operator{L}_S\right)$ supplies an orthogonal linear mapping in Euclidean space $\Re _3$, so the reference parallel transport of tangent vectors along the circle $C _s$ is mapped to the corresponding transport of their images along $C_{\lambda}$.

Fig.~\ref{fig:ParallelTransportXYplane} illustrates the reference transport of initial tangent vector $\vector{\widetilde{\Psi}_0} $ along a closed circular path $C_s$.
The current tangent plane $\Lambda _D(\varphi )$ is spanned by natural two basis vectors $\vector{e}_\varphi (\varphi), \vector{e}_\theta (\varphi)$ corresponding to azimuthal and polar coordinate angles $\varphi$ and $\theta$ respectively.
Angle
\begin{eqnarray}
    \label{eq:geometricPhaseBetaG}
	\widetilde{\beta }_{\varphi}=-(2\pi-\beta) /(2\pi)\cdot \varphi =-\sin{\lambda}/(2\pi)\cdot \varphi
\end{eqnarray}
measures the partial rotation of vector $\widetilde{\Psi}$ in basis $\vector{e}_\varphi, \vector{e}_\theta$ accumulated by  $\widetilde{\Psi}$ in point $\vector{e}_R (\varphi)$ upon  its parallel transport \cite{Arnold_1978} (also Eq.~(\ref{eq:AngleA12}) in \ref{appendix:EvolutionA2}). The minus sign here corresponds to measuring the rotation angles $\widetilde{\beta}_{\varphi}$  in the clockwise direction from the vector's initial position, as viewed from the positive direction of the radial axis $\vector{e}_R$ (see also Fig.~231 of book \cite{Arnold_1978} on page 302). Since the geometric factor (\ref{eq:betaRelationTolambda}) is  measured in the counterclockwise direction (the reference positive direction in Riemannian geometry \cite{Arnold_1978}) the total  rotation angle $|\widetilde{\beta }_{2\pi}|$ is complementary to $\beta$: $|\widetilde{\beta }_{2\pi}|+\beta =2\pi$.

\subsection{Parallel transport along the actual circle $C_{\lambda}$ }
\label{sect:ActualTransport32}

Relation~(\ref{eq:trajectoriesRelationViaRotationGroupGenerator})  implies that parallel transport of a tangent vectors along $C_s$ is orthogonally mapped onto $C_{\lambda}$ and vice versa.
Thus, the adiabatic passage procedure can be performed in three consecutive steps:

(i) The initial vectors corresponding to the trajectory $C_{\lambda}$ are rotated around the $S$-axis by angle $-\lambda$:
\begin{eqnarray}
    \label{eq:radiusVectorFirstStepSTIRAP}
    \vector{e}_{R_0} = \vector{e}_Q  \to \tilde{\vector{e}}_{R_0}= \exp{\left(-\lambda\operator{L}_S\right)}\vector{e}_Q,\\
    \label{eq:stateVectorFirstStepSTIRAP}
    \vector{\Psi}_0 = \vector{e}_P \to \tilde{\vector{\Psi}}_0 = \exp{\left(-\lambda\operator{L}_S\right)}\vector{e}_P;
\end{eqnarray}

(ii) The bright vector $\tilde{\vector{e}}_{R_0}$ revolves around the $P$-axis by azimuthal angle $\varphi = 2\pi$, describing the trajectory $C_s$:
\begin{eqnarray}
    \label{eq:radiusVectorSecondStepSTIRAP}
    \tilde{\vector{e}}_{R_0} \to  \tilde{\vector{e}}_R(\varphi) = \exp{\left(\varphi\operator{L}_P\right)}\tilde{\vector{e}}_{R_0}
\end{eqnarray}

As a result of parallel transport along $C_s$, the tangent vector $\tilde{\vector{\Psi}}_0$ rotates by $\widetilde{\beta }_{\varphi}$ (\ref{eq:geometricPhaseBetaG}) around the radial axis $\tilde{\vector{e}}_R(\varphi)$ in the current tangent plane $\{\vector{e}_\theta(\varphi),\vector{e}_\varphi(\varphi)\}$:
\begin{eqnarray}
    \label{eq:stateVectorSecondStepSTIRAP}
    \tilde{\vector{\Psi}}_0 \to \tilde{\vector{\Psi}}_\varphi
    =
    \exp{\left(\widetilde{\beta }_{\varphi}\operator{L}_{\widetilde{R}(\varphi)}\right)}
    \tilde{\vector{\Psi}}_0
\end{eqnarray}

(iii) In the final step, we must return to the original trajectroy $C_{\lambda}$, rotating the vectors (\ref{eq:radiusVectorSecondStepSTIRAP},\ref{eq:stateVectorSecondStepSTIRAP}) around the $S$-axis by $\lambda$:
\begin{eqnarray}
    \label{eq:radiusVectorThirdStepSTIRAP}
    \tilde{\vector{e}}_R(\varphi) \to \vector{e}_R(\varphi)
    =
    \exp{\left(\lambda\operator{L}_S\right)}
    \exp{\left(\varphi\operator{L}_P\right)}
    \exp{\left(-\lambda\operator{L}_S\right)}
    \vector{e}_Q
    \\
     \label{eq:stateVectorThirdStepSTIRAP}
    \tilde{\vector{\Psi}}_\varphi \to \vector{\Psi}_\varphi
    =
    \exp{\left(\lambda\operator{L}_S\right)}
    \exp{\left(\widetilde{\beta }_{\varphi}\operator{L}_{\widetilde{R}(\varphi)}\right)}
    \exp{\left(\varphi\operator{L}_P\right)}
    \exp{\left(-\lambda\operator{L}_S\right)}
    \vector{e}_P;
\end{eqnarray}
The expression (\ref{eq:radiusVectorThirdStepSTIRAP}) describes evolution of the reduced laser Rabi frequencies  $\widetilde{\Omega} _j$ (\ref{eq:E13}) along the Rabi  trajectory $C_{\lambda}$ (Fig.~\ref{fig:ParallelTransportSTIRAP}):
\begin{eqnarray}
    \label{eq:RabiQE24}
    \tilde{\Omega}_Q(\lambda, \varphi) &=& \cos^{2}{\lambda}\cos{\varphi} + \sin^{2}{\lambda}
    \\
    \label{eq:RabiSE25}
    \tilde{\Omega}_S(\lambda, \varphi)  &=& \cos{\lambda}\sin{\varphi}
    \\
    \label{eq:RabiPE26}
    \tilde{\Omega}_P(\lambda, \varphi)  &=& \cos{\lambda}\sin{\lambda}\left(1-\cos{\varphi}\right)
\end{eqnarray}
The angular parameter $\lambda$ is fixed by the chosen value of $\beta$ (\ref{eq:betaRelationTolambda}), according to the required fractional STIRAP mixing angle (\ref{eq:StateAfterFractionalSTIRAP}), while the azimuth $\varphi\in\left[0,2\pi\right]$ plays the role of the temporal parameter.
The laser Rabi frequencies $\Omega_i=\overline{\Omega} \tilde{\Omega}_i $ derived this way satisfy the Dykhne-Davis-Pechukas optimal adiabaticity criterion \cite{Dykhne1960, DavisPechukas1976}, when throughout the experiment
\begin{eqnarray}
  \label{eq:DykhnePE27}
    \Omega^{2}_{eff}(\varphi) = \Omega^2_P + \Omega^2_Q + \Omega^2_S =\overline{\Omega}^2 = const.,
\end{eqnarray}
ensuring minimal population exchange between the dark and bright subspaces.

The dependence of the dark state-vector $\vector{\Psi} = \{a_Q,a_S,a_P\} \to \ket{\Psi} = a_P\ket{1} + a_S\ket{3} + a_Q\ket{4}$ is obtained by performing the successive rotations in (\ref{eq:stateVectorThirdStepSTIRAP}), yielding:

\begin{eqnarray}
   \label{eq:AmplitudePE28}
    a_P
    =
    &\sin^{2}{\lambda}\cos{\left(\widetilde{\beta }_{\varphi}\right)}\cos{\varphi}
    -
    \sin{\lambda}\sin{\left(\widetilde{\beta }_{\varphi}\right)}\sin{\varphi}
    + \cos^{2}{\lambda}\cos{\left(\widetilde{\beta }_{\varphi}\right)}
    \\
    \label{eq:AmplitudeQE29}
    a_Q
    =
    &-\cos{\lambda}\sin{\lambda}\cos{\left(\widetilde{\beta }_{\varphi}\right)}\cos{\varphi}
   +\cos{\lambda}\sin{\left(\widetilde{\beta }_{\varphi}\right)}\sin{\varphi} \nonumber\\
    &+\cos{\lambda}\sin{\lambda}\cos{\left(\widetilde{\beta }_{\varphi}\right)}
    \\
    \label{eq:AmplitudeSE30}
    a_S
    =
    &-\sin{\lambda}\cos{\left(\widetilde{\beta }_{\varphi}\right)}\sin{\varphi}
   -\sin{\left(\widetilde{\beta }_{\varphi}\right)}\cos{\varphi}
\end{eqnarray}
Here angle $\widetilde{\beta }_{\varphi}$ is determined by Eq.(17).

It is easy to verify that the above equations describe the passage of initial state $\ket{1}$ at $\varphi=0$ to the superposition state (\ref{eq:StateAfterFractionalSTIRAP}) at $\varphi=2\pi$.

\section{Illustrations by numerical simulations  and discussion}
\label{sect:Examples}

\begin{figure}[ht]
    \centering
    \includegraphics[width=0.6\textwidth]{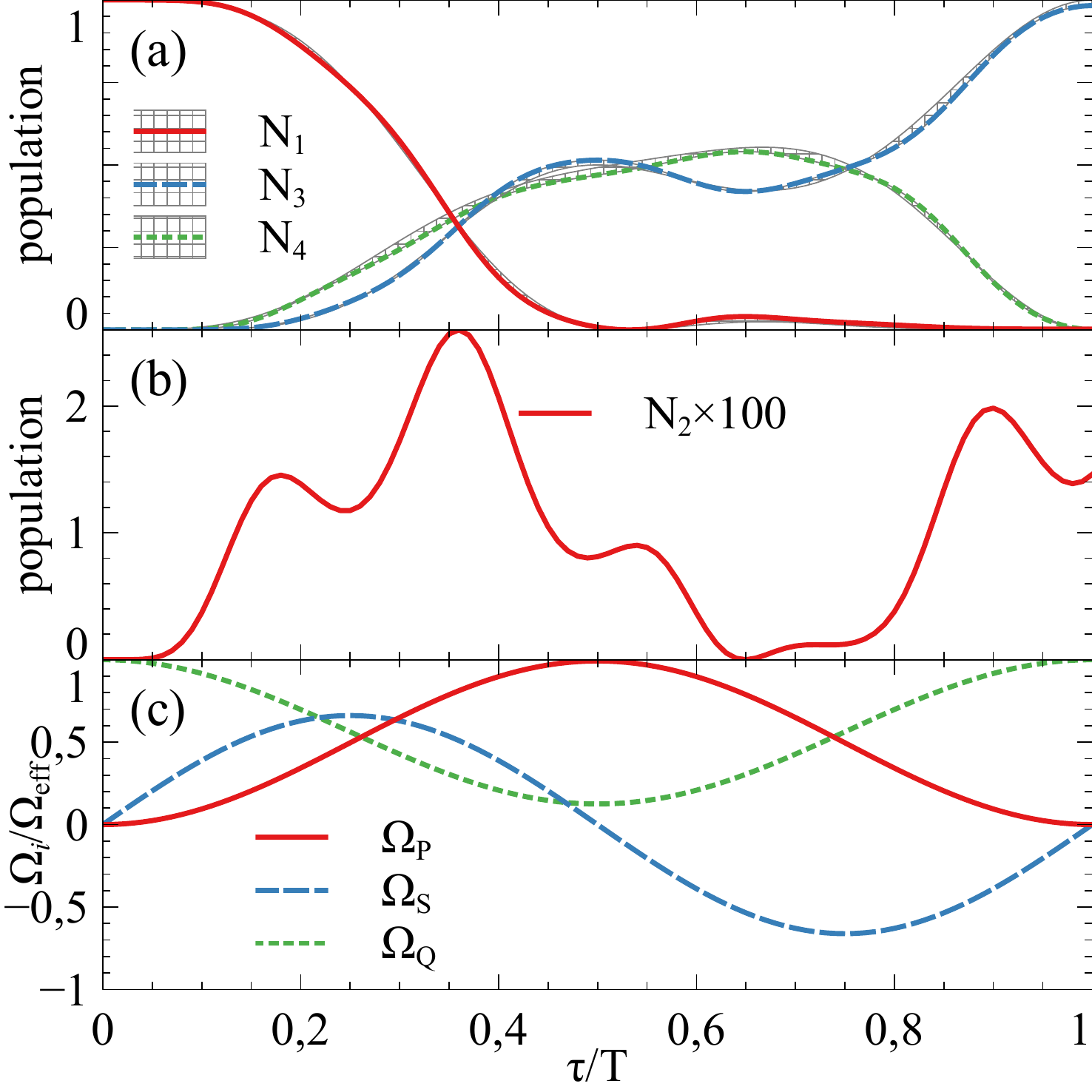}
    \caption{
      "Full" STIRAP ($\beta=\pi/2$) for adiabaticity parameter $\wp=30$ with harmonic laser pulses (\ref{eq:OmegaQv2}-\ref{eq:OmegaPv2}) and zero single-photon detuning ($\Delta =0$).
      (a) Population dynamics; highlighted areas represent deviation from the analytical results (\ref{eq:AmplitudePE28}-\ref{eq:AmplitudeSE30}).
      (b) Population of the excited state $\ket{2}$, magnified $100$ times for convenience.
      (c) The corresponding sequence of harmonic laser pulses.
    }
    \label{fig:FullStirapHarmonicPulses}
\end{figure}

\begin{figure}[ht]
    \centering
    \includegraphics[width=0.8\textwidth]{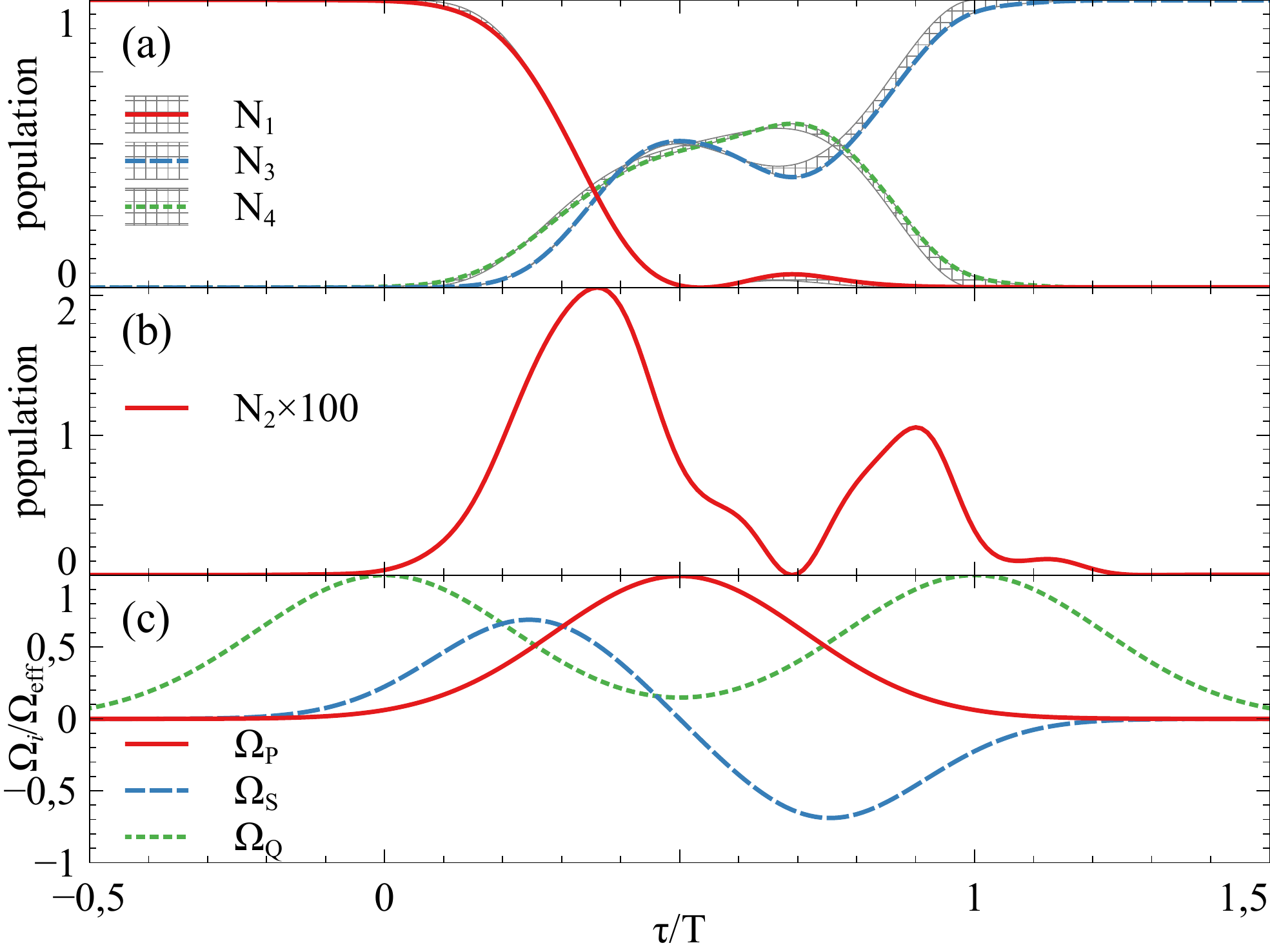}
    \caption{
     The same as in Fig.~\ref{fig:FullStirapHarmonicPulses} for the "full" STIRAP ($\beta=\pi/2$), but with  Gaussian-shaped laser pulses (\ref{eq:OmegaQgaussian}-\ref{eq:OmegaPgaussian})}.
    \label{fig:FullStirapGaussianPulses}
\end{figure}

\begin{figure}[ht]
    \centering
    \includegraphics[width=0.6\textwidth]{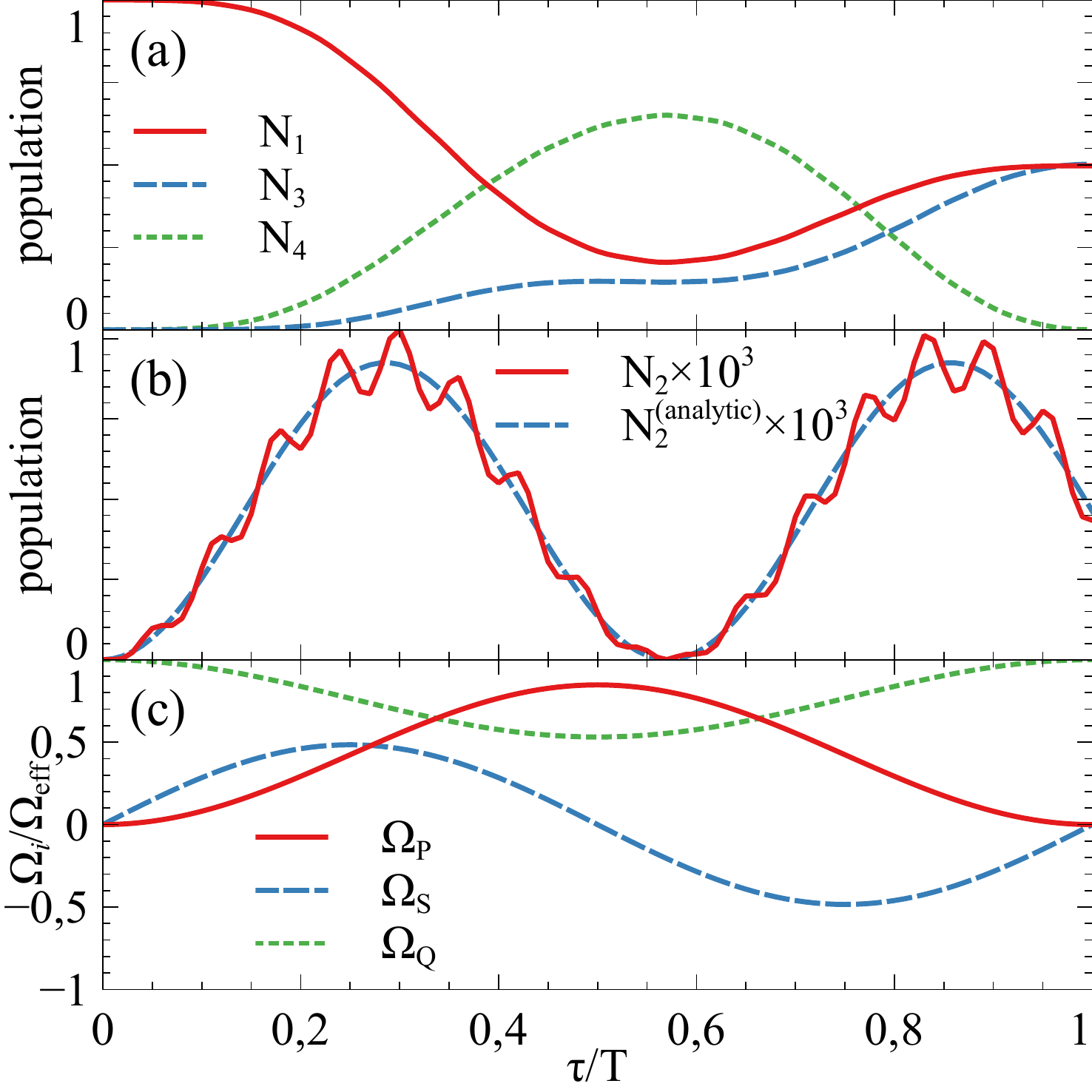}
    \caption{
The same as in Fig.~\ref{fig:FullStirapHarmonicPulses} for the "half" STIRAP ($\beta=\pi/4$) and  adiabaticity parameter $\wp=100$.  The numerical results in frame (a) are practically indistinguishable from the analytical ones. The dashed curve in a frame (b) is  plotted via expression~(\ref{eq:AmplitudeC_2}): $N_2=|C_2|^2$.
    }
    \label{fig:HalfStirapHarmonicPulses}
\end{figure}

\begin{figure}[ht]
    \centering
    \includegraphics[width=0.6\textwidth]{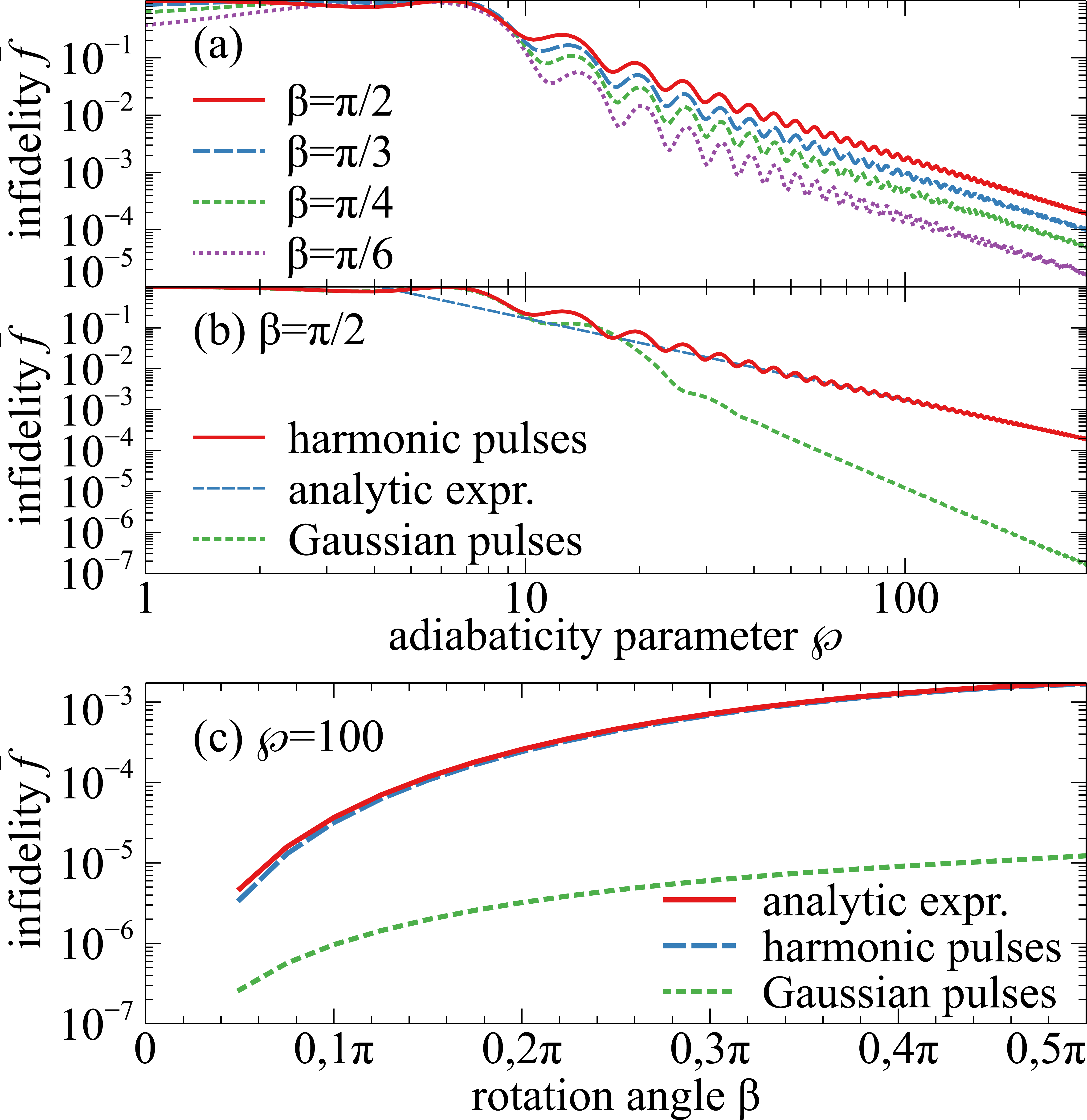}
    \caption{
      (a)  STIRAP infidelities (\ref{eq:InfidelityE38}) for various rotation angles $\beta$ as a function of the adiabaticity parameter $\wp$ in the case of harmonic pulses.
      (b) Infidelity of the "full" STIRAP ($\beta=\pi/2$) as a function of the adiabaticity parameter $\wp$, comparison between numerical results for the harmonic laser pulses and their Gaussian approximation.
      (c) STIRAP infidelity as a function of the rotation angle $\beta$ for the harmonic laser pulses and the Gaussian approximations.
      In both (b) and (c), the analytical infidelity (\ref{eq:InfidelityE40}) of harmonic pulse sequence is also shown for reference.
    }
    \label{fig:Infidelities}
\end{figure}

When describing analytical laser pulses (\ref{eq:RabiQE24}-\ref{eq:RabiPE26}), we used dimensionless temporal parametrization.
Let us introduce a new set of parameters that explicitly contain the STIRAP time scale, namely $\alpha = \frac{2}{\pi}\lambda, \alpha\in[0,1]$ and $\tau = \frac{\varphi}{2\pi}T, \tau\in[0,T]$, where $T$ denotes the time duration of the laser pulses.
Taking into account that the laser Rabi frequencies are proportional to the reduced Rabi frequencies $\tilde{\Omega}_i$, these substitutions allow us to rewrite the expressions (\ref{eq:RabiQE24}-\ref{eq:RabiPE26}) in the following form:
\begin{eqnarray}
    \label{eq:OmegaQv2}
    \Omega_Q(\alpha, \tau)
    &=
    \Omega _{eff}\left(\cos^{2}\of{\frac{\pi}{2}\alpha}\cos\of{2\pi\frac{\tau}{T}} + \sin^{2}\of{\frac{\pi}{2}\alpha}\right)
    \\
    \label{eq:OmegaSv2}
    \Omega_S(\alpha, \tau)
    &=
    \Omega _{eff}\cos\of{\frac{\pi}{2}\alpha}\sin\of{2\pi\frac{\tau}{T}}
    \\
    \label{eq:OmegaPv2}
    \Omega_P(\alpha, \tau)
    &=
    \Omega _{eff}\cos\of{\frac{\pi}{2}\alpha}\sin\of{\frac{\pi}{2}\alpha}\left(1-\cos\of{2\pi\frac{\tau}{T}}\right)
\end{eqnarray}
The parameter $\alpha$ is determined by the desired rotation angle $\beta $~(\ref{eq:E1})
while the adiabaticity parameter $\wp $ ~(\ref{eq:E11}) is proportional to the product $T\Omega_{eff}$:
\begin{eqnarray}
\label{eq:ParametersE34}
    \alpha &= \frac{2}{\pi}\arcsin\of{1-\beta /2\pi};  \quad
\wp = \frac{1}{2}T\Omega_{eff}
\end{eqnarray}

\subsection{STIRAP simulations}
\label{sect:Simulations41}

In this subsection we consider examples of full STIRAP ($\beta =\pi/2$) (Fig.~\ref{fig:FullStirapHarmonicPulses},\ref{fig:FullStirapGaussianPulses}) and half-STIRAP ($\beta =\pi/4$) (Fig.~\ref{fig:HalfStirapHarmonicPulses}) when the single-photon detuning is zero: $\Delta =0$. 
The data presented in Figs.~\ref{fig:FullStirapHarmonicPulses}-\ref{fig:HalfStirapHarmonicPulses} were obtained by numerically solving the Schrödinger equation (\ref{eq:E2}) (bold lines). 
When distinguishable, the deviations from the analytical expressions (\ref{eq:AmplitudePE28})-(\ref{eq:AmplitudeSE30}) are highlighted with thin lines in frame (a) of each figure.

The harmonic laser pulses (\ref{eq:OmegaQv2}-\ref{eq:OmegaPv2}) for $1/2\leq\alpha\leq1$ have the important advantage of being reasonably well approximated by a set of Gaussian laser pulses:
\begin{eqnarray}
    \label{eq:OmegaQgaussian}
    \Omega^{(G)}_Q(\tau)
    &\propto
    \exp\of{-\frac{\of{\tau/T}^2}{\sigma_Q^2}} + \exp\of{-\frac{\of{\tau/T-1}^2}{\sigma_Q^2}}
    \\
    \label{eq:OmegaSgaussian}
    \Omega^{(G)}_S(\tau)
    &\propto
    \exp\of{-\frac{\of{\tau/T-1/4}^2}{\sigma_S^2}} - \exp\of{-\frac{\of{\tau/T-3/4}^2}{\sigma_S^2}}
    \\
    \label{eq:OmegaPgaussian}
    \Omega^{(G)}_P(\tau)
    &\propto
    \exp\of{-\frac{\of{\tau/T-1/2}^2}{\sigma_P^2}}
\end{eqnarray}
The proportionality coefficients and widths $\sigma_{Q,S,P}$ are fitting parameters.
Optimal values of the fitting parameters are obtained by taking into account three aspects: (1) the Gaussian functions must preset a reasonable approximation for the harmonic pulses; (2) The pulse sequence must yield the same rotation angle $\beta$ (\ref{eq:E12}) as the harmonic pulses sequence; (3) optimal adiabaticity requires that deviations from $\Omega_{eff}=\textnormal{const.}$ are minimal.
Figure \ref{fig:FullStirapGaussianPulses} illustrates the evolution of the level populations for a "full" STIRAP driven by  Gaussian-shaped laser pulse trains.

\subsection{Discussion}
\label{sect:Discussion42}

The numerical data exhibited  in Figs.~\ref{fig:FullStirapHarmonicPulses}-\ref{fig:HalfStirapHarmonicPulses} for the populations of the ground states $\ket{j}$ demonstrate good agreement with the analytical results obtained within the framework of the geometric approach. Here we discuss the issue of the efficiency of STRAP and give an estimate of the population fraction transferred to the unwanted excited state $\ket{2}$.

Commonly, the fidelity \cite{Shore2017} of a coherent control method is defined via the overlap between the states required and actually produced.
For high-fidelity methods it more convenient to introduce measure of infidelity $\bar{f}$ as the method's deviation from the perfect fidelity of $1$:
\begin{eqnarray}
 \label{eq:InfidelityE38}
    \bar{f}=1-\abs{\braket{\Psi_{\textnormal{ideal}}}{\Psi}}^2.
\end{eqnarray}
In our case, formula (\ref{eq:InfidelityE38}) reduces to $\bar{f}=N_2$, i.e. to the population lost in the subspace of dark states due to the non-adiabatic mixing between bright and dark states.

Figure~\ref{fig:Infidelities}  demonstrates the dependence of infidelity $\bar{f}$ on the adiabaticity parameter $\wp $ and the geometrical factor $\beta$ for different profiles of laser pulses. As in the case of the reference STIRAP in $\Lambda$-configurations \cite{Vitanov2017}, the oscillations of the infidelity parameter can be attributed to Rabi oscillations in a pair of strongly coupled bright and excited states (Fig.~\ref{fig:tripod}b). Unlike STIRAP in $\Lambda$-systems (\cite{Vitanov2013}), these oscillations fade with increasing pulse area and do not limit the accuracy of adiabatic passage as can be judged from Fig.~\ref{fig:Infidelities}a.

These numerically obtained findings are well justified by analytical results derived in \ref{appendix:InfidelityC} for probability amplitude  $C_2$ of excited state $\ket{2}$
\begin{eqnarray}
 \label{eq:AmplitudeC_2}
  C_{2}\cong \frac{2M_{DB}}{\hbar \Omega _{eff}}=\im\frac{ \sqrt{(4\pi -\beta )\beta}}{\wp}\sin\of{(1-\beta/2\pi)\varphi}
\end{eqnarray}
and infidelity parameter
\begin{eqnarray}
 \label{eq:InfidelityE40}
    \bar{f}=|C_{2}|^2\cong \frac{4\pi -\beta }{\wp^2}\beta   \sin^{2}\of{\beta}
\end{eqnarray}
which are regulated with the non-adiabatic coupling between dark and bright states. In the strong adiabaticity limit, the amplitude $C_2$ instantly follows a relatively slow change in the corresponding nonadiabaticity matrix element $M_{DB}$ (\ref{eq:MatrixElementC3}) and turns out to be weakly subject to fast Rabi oscillations with frequency $\Omega _{eff}$.

Another advantage of geometric pulses (\ref{eq:OmegaQv2})-(\ref{eq:OmegaPv2}) is associated with the possibility of their sufficiently accurate approximation by Gaussian pulse sequences (\ref{eq:OmegaQgaussian})-(\ref{eq:OmegaPgaussian}), convenient for practical applications. As is typical for the general situation, the exponential smoothing of the moments of switching on and off of the interaction of light with matter leads to significantly higher accuracy of STRAP processes.

\section{Conclusion}
\label{sect:Conclusion}
In this paper, we have analyzed the process of adiabatic passage in N-pod atomic systems, focusing on tripod STIRAP.
The common structure of the energy level diagrams and excitation schemes (Fig.~\ref{fig:tripod}) makes it possible to study adiabatic processes in these systems from the unified standpoint of Riemannian geometry.
The stable states subspace is decomposed into the subspaces of adiabatic dark states, decoupled from the laser control fields, and the single bright state responsible for the excitation process.
A geometrical analogy is proposed for the N-pod system dynamics, where the bright state is associated with the N-dimensional radius-vector in the parameter space of laser Rabi frequencies.
The bright state describes a certain curve (Rabi curve) on an (N-1)-dimensional unit sphere, which can be viewed as a generalization of the Bloch sphere.
The dark states lie in a plane adjoining the Bloch sphere, where their dynamics is reduced to Riemannian parallel transport of the "dark" tangent plane along the Rabi curve on the Bloch sphere.

Using the geometrical approach to the analysis of tripod STIRAP, we have demonstrated how the parameters of the adiabatic passage can be optimized by choosing Rabi curves in the form of circles on the Bloch sphere.
The laser pulse sequences are constructed in such a way that the dark states subspace both initially and at the end of an experiment is aligned with the subspace of the same two tripod ground states, allowing one to use our result in designing reversible rotation quantum gates (\ref{eq:E1}).
An important advantage of the proposed pulse sequences is the blocking of oscillations in STIRAP fidelity arising from Rabi oscillations in the pair of coupled states.
As a consequence, such fidelity-limiting oscillations are not observed, allowing for high-fidelity adiabatic passage.

\section{Acknowledgments}
This work was supported  by Latvian Council of Science grant No. LZP-2019/1-0280.

\appendix

\section{
    Nonadiabatic transitions as Riemannian parallel transport
\label{appendix:NonadiabaticA}
}
Nonadiabatic evolution of the two-dimensional subspace $\Lambda_D$ of degenerate dark states leads to a unitary transformation of the dark  basis states $\ket{D_k}$.
In the three-dimensional parameter space
$\Re_ 3$  of laser Rabi frequencies defined in Sect.~\ref{sect:Bright21}, the single bright state is associated with unit radius vector $\vector{e}_R\propto \{\Omega_Q,\Omega_S,\Omega_P\}$.
Upon parametric evolution of the Rabi frequencies $\Omega_j\of{s}, s\in[0,\infty)$, the vector $\vector{e}_R\of{s}$ follows a parametric curve $\Im _1$ on the unit Bloch sphere, while the two-dimension flat  subspace $\Lambda_D\of{s}$ is always tangent (orthogonal) to $\vector{e}_R\of{s}$.
The parameter $s$ can always be redefined via the natural (or arc-length) parametrization so that its current value gives the length $l\of{s}\equiv{s}$ of the curve \cite{Kreyszig1991}.

\subsection{Reference basis sets for tangent planes of a curve $\Im _1$
\label{appendix:ReferenceA1}}

For the two-dimensional tangent dark spaces $\Lambda_D\of{s}$, one can choose a convenient basis set of dark states $\vector{e}_\varphi\of{s},\vector{e}_\theta\of{s}$.
The first of them is the unit tangent vector to the curve at point $\vector{e}_R\of{s}$, and its cross product with $\vector{e}_R\of{s}$ determines the other dark state (Fig.~\ref{fig:ParallelTransportXYplane}):
\begin{eqnarray}
    \label{eq:BasisVectorPhi}
    \vector{e}_\varphi\of{s} &= \frac{d}{ds}\vector{e}_R\of{s}
    \\
    \label{eq:BasisVectorTheta}
    \vector{e}_\theta\of{s} &= \vector{e}_\varphi\of{s}\times\vector{e}_R\of{s}
\end{eqnarray}

The standard framework of differential geometry defines another basis set for the trajectory $\Im _s$, the so-called Frenet frame or TNB frame \cite{Kreyszig1991},
\begin{eqnarray}
    \label{eq:TNB_basis}
    \vector{T}_s = \vector{e}_\varphi\of{s};
    \quad
    \frac{d}{ds}\vector{T}_s = \kappa\vector{N}_s;
    \quad
    \vector{B}_s = \vector{N}_s\times\vector{T}_s.
\end{eqnarray}
According to the Frenet-Serret formulas \cite{Kreyszig1991}, the normal $\vector{N}_s$ is related to the derivative of $\vector{T}_s$ via curvature $\kappa$ of the curve, and the binormal $\vector{B}_s$ determines torsion $\tau$ of the curve, $d\vector{B}_s/ds = -\tau\vector{N}_s$.

The tangent $\vector{T}_s$ and the normal $\vector{N}_s$  vectors define the osculating plane $\Pi_s$ at point $\vector{e}_R\of{s}$.
For a curve $\Im _1$ lying on the unit sphere, this plane cuts out of the unit sphere a osculating circle $C_s$ of radius $r_s=1/\kappa$. The circle $C_s$ and the curve $\Im _1$ coincide in the second order in $ds$, i.e. they are weakly distinguishable in a small neighborhood of the current parameter  $s$.

If for the selected point $\vector{e}_R\of{s}$, we take the binormal $\vector{B}_s$ as the direction of the local P-axis,  i.e. $\vector{e}_p(s)=\vector{B}_s$, then the above current configuration of all basis vectors, the plane $\Pi_s$, and the circle $C_s$ will coincide with the notation given in  Fig.~\ref{fig:ParallelTransportXYplane}.
Formulas, presented in  Eq.~(\ref{eq:TNB_basis}), result in  an important equality
\begin{eqnarray}
    \label{eq:importantRelationAngleTheta}
   \frac{d}{ds} \vector{e}_R(s) = \frac{1}{\sin{\theta_s}}\vector{e}_z\times\vector{e}_R\of{s},
    \quad
    \sin{\theta_s} = r_s = 1/\kappa,
\end{eqnarray}
where $\theta_s$ is the angle between the radius vector $\vector{e}_R$  and $P$-axis $\vector{e}_p$.
The expression (\ref{eq:importantRelationAngleTheta}) corresponds to rotation of the bright state $\vector{e}_R$ around $P$-axis with a local angular velocity $\kappa$.
As a direct consequence of relations (\ref{eq:TNB_basis}), the tangent vector $\vector{e}_\varphi\of{s}$ rotates at the same velocity,
\begin{eqnarray}
    \frac{d}{ds}\vector{e}_\varphi\of{s} &= \kappa\vector{N}_s = \kappa\vector{e}_p\times\vector{e}_\varphi\of{s}.
\end{eqnarray}
Thus, the binormal axis determines the local rotation direction and velocity $d\varphi/ds = \kappa$ of both bright and dark basis vectors, and the tangent plane $\Lambda_D\of{s}$ itself.
After passing the curve segment $\vector{e}_\varphi\of{s}ds$, the plane $\Lambda_D\of{s}$ rotates around $\vector{e}_p$ by angle $d\varphi = \kappa ds = ds/r_s$, which can be formally rewritten via $P$-axis rotation group generator $\operator{L}_P$ as
\begin{eqnarray}
    \label{eq:TangentPlaneEvolution}
    \Lambda_D\of{s+ds} &= \exp\of{\kappa ds \operator{L}_P}\Lambda_D\of{s}.
\end{eqnarray}

\subsection{Evolution of dark state-vector driven by nonadiabatic coupling
\label{appendix:EvolutionA2}}
Consider a state-vector $\vector{\Psi}\of{s}$ residing in the dark tangent space $\Lambda_D\of{s}$,
\begin{eqnarray}
    \label{eq:StateVectorInTangentBasis}
    \vector{\Psi}\of{s} &= \cos{\beta_s}\vector{e}_\theta\of{s} + \sin{\beta_s}\vector{e}_\varphi\of{s},
\end{eqnarray}
with $\beta_s$ representing a mixing angle.
Let us determine from the Schrodinger equation how $\beta_s$ changes when the parameter $s$ changes by $ds$.
The nonadiabatic coupling operator in a set of degenerate dark states has the form \cite{Wilczek1984, Aharonov1987, Unanyan_1999}
\begin{eqnarray}
    H_{\alpha\alpha^\prime} &= \im\hbar\frac{ds}{dt}\melem{\vector{e}_\alpha\of{s}}{\frac{d}{ds}}{\vector{e}_{\alpha^\prime}\of{s}},
\end{eqnarray}
where $\alpha(\alpha^\prime) =\vartheta,\varphi$. Orthonormality of the basis vectors implies vanishing diagonal elements $H_{\alpha\alpha}=0$.
According to (\ref{eq:TNB_basis}), the off-diagonal elements become:
\begin{eqnarray}
    H_{\theta\varphi} &= \im\hbar\frac{ds}{dt}\kappa\braket{\vector{e}_\alpha\of{s}}{\vector{N_s}}.
\end{eqnarray}
For the current osculating circle  $C_s$ (Fig.~\ref{fig:ParallelTransportXYplane}), this yields:
\begin{eqnarray}
    H_{\theta\varphi} &= \im\hbar\kappa\cos{\theta _s}\frac{ds}{dt}.
\end{eqnarray}
The resulting non-adiabatic evolution equation
\begin{eqnarray}
    \label{eq:statevectorEvolutionRotation}
    \im\hbar\frac{d}{dt}\vector{\Psi}\of{s}
    &=
    -\im\hbar\kappa\cos{\theta _s}\cdot\frac{ds}{dt}\left(\matrix{
        0   &   1 \cr
        -1  &   0 \cr
    }\right)\vector{\Psi}\of{s}
\end{eqnarray}
describes rotation of the state-vector in the tangent plane $\Lambda_D\of{s}$ around the current radial axis $\vector{e}_R\of{s}$.
Inserting (\ref{eq:StateVectorInTangentBasis}), we obtain the following relation for evolution of the mixing angle $\beta_s$:
\begin{eqnarray}
\label{eq:AngleA12}
    d\beta_s = -\cos{\theta_s}\frac{ds}{r_s} = -\cos{\theta _s}d\varphi.
\end{eqnarray}
Evolution of the state-vector (\ref{eq:statevectorEvolutionRotation}) can also be expressed as:
\begin{eqnarray}
    \label{eq:StateVectorEvolutionFromQuantumMechanics}
    \vector{\Psi}\of{s+ds} = \exp\of{-\cos{\theta _s}d\varphi\operator{L}_{R_s}}\vector{\Psi}\of{s}.
\end{eqnarray}
Here $\operator{L}_{R_s}$ is the generator of the rotation group about $\vector{e}_R\of{s}$ axis - i.e. around the bright state.
The negative sign corresponds to clockwise rotation as viewed from the positive direction of the radial axis.

\subsection{Evolution of dark state-vector under Riemannian parallel transport
\label{appendix:EvolutionA3}}
The relation (\ref{eq:StateVectorEvolutionFromQuantumMechanics}) can also be obtained in the framework of Riemannian geometry where it emerges as a consequence of the parallel transport of a tangent vector along curved surface of a sphere.
Consider parallel transport $\vector{\Psi}\of{s} \to \vector{\Psi}\of{s+ds}$ of a tangent vector $\vector{\Psi}\of{s} \in \Lambda_D\of{s}$ along a small rectilinear segment $d\vector{e}_R\of{s} = \vector{e}_\varphi\of{d}ds$ of the curve $C_s$.
The transport procedure is carried out as a two-step process \cite{Kreyszig1991,Arnold_1978}:
(i) free translation of $\vector{\Psi}\of{s}$ to the point $\vector{e}_R\of{s+ds}$,
and (ii) projection of $\vector{\Psi}\of{s}$ onto the new tangent plane $\Lambda_D\of{s+ds}$.

The two tangential planes $\Lambda_D\of{s+ds}$ and $\Lambda_D\of{s}$ are related via rotation (\ref{eq:TangentPlaneEvolution}), implying that the vector $\vector{\Psi}\of{s}$ in the new plane $\Lambda_D\of{s+ds}$ appears to be rotated around $P$-axis by $-d\varphi = -\kappa ds$:
\begin{eqnarray}
    \label{eq:VectorParallelTransportStep1}
    \vector{\Psi}_{ds}\of{s} &= \exp\of{-d\varphi \operator{L}_P}\vector{\Psi}\of{s}.
\end{eqnarray}
The second step of projecting the vector $\vector{\Psi}_{ds}\of{s}$ onto $\Lambda_D\of{s+ds}$ in the first approximation in $ds$ is reduced to projection $\operator{\Xi}_D$ onto the original plane $\Lambda_D\of{s}$:
\begin{eqnarray}
    \label{eq:VectorParallelTransportStep2}
    \vector{\Psi}\of{s+ds} &= \operator{\Xi}_D\exp\of{-d\varphi \operator{L}_P}\vector{\Psi}\of{s}.
\end{eqnarray}
The projection operator $\operator{\Xi}_D$ becomes identity operator for vectors parallel to the tangent plane $\Lambda_D\of{s}$, but it vanishes vectors orthogonal to this plane.
An infinitesimal rotation $\exp\of{-d\varphi\operator{L}_P}$ can be implemented as a sequence of two auxiliary rotations by decomposing the vector
\begin{eqnarray}
    \label{eq:InfinitesimalRotationVectorDecomposition}
    \vector{\Omega}
    =
    d\varphi \vector{e}_P\of{s}
    =
    d\varphi\cos{\theta_s}\vector{e}_{R_{s}}
    + d\varphi\sin{\theta _s}\vector{e}_{D_{s}}
\end{eqnarray}
into projections along the radial axis $\vector{e}_{R_{s}}$ (first term) and projection onto the plane $\Lambda_D\of{s}$ (second term).
The infinitesimal rotation $\exp\of{-d\varphi\operator{L}_P}$ is reduced \cite{Arnold_1978} to rotation around the radial axis $\vector{e}_{R_{s}}$ by angle $d\varphi\cos{\theta _s}$ and a subsequent rotation around $\vector{e}_{D_{s}}$ by angle $d\varphi\sin{\theta _s}$, allowing us to rewrite (\ref{eq:VectorParallelTransportStep2}) as
\begin{eqnarray}
    \label{eq:VectorParallelTransportStep2v2}
    \vector{\Psi}\of{s+ds} &= \operator{\Xi}_D\exp\of{-d\varphi\sin{\theta _s}\operator{L}_{D_s}}\exp\of{-d\varphi\cos{\theta _s}\operator{L}_{R_s}}\vector{\Psi}\of{s}.
\end{eqnarray}
It is easy to show that (\ref{eq:VectorParallelTransportStep2v2}) is equivalent to (\ref{eq:StateVectorEvolutionFromQuantumMechanics}).
The rightmost operator is identical to one in relation (\ref{eq:StateVectorEvolutionFromQuantumMechanics}), and leaves the vector in the plane $\Lambda_D\of{s}$.
The second operator adds to the vector an infinitesimal vector $d\vector{\Psi}$, which is orthogonal to the plane $\Lambda_D\of{s}$, and therefore vanishes upon projection $\operator{\Xi}_D$, while the rotated vector remains unchanged as it already belong to the plane $\Lambda_D\of{s}$.

\section{
    Contour integral for geometric phase $\beta$
    \label{appendix:BerryPhase}
}
\begin{figure}[h]
    \centering
    \includegraphics[width=0.50\textwidth]{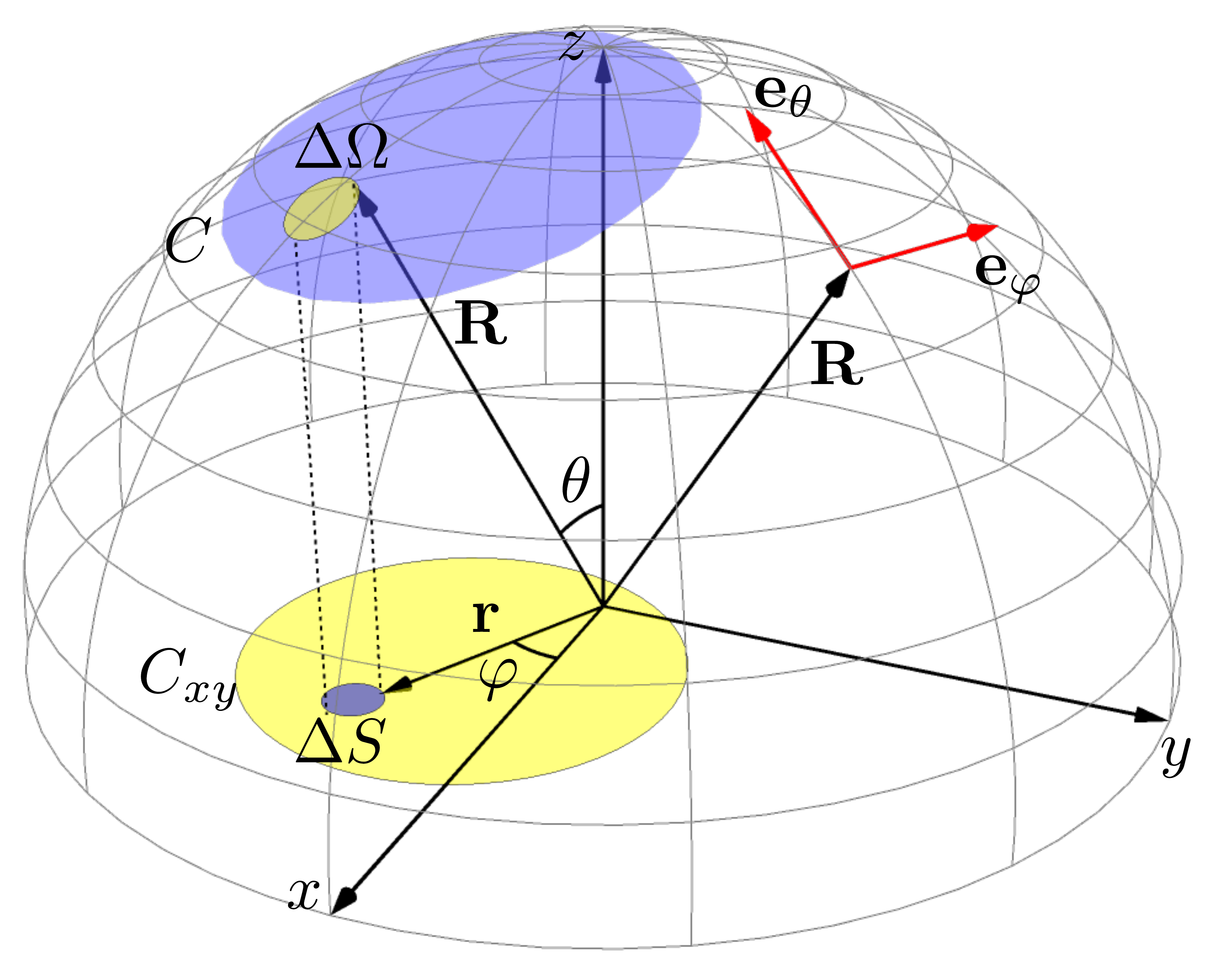}
    \caption{
        Projection onto the coordinate plane $(x,y)$ of a spherical surface spanned by the contour $C$.
    }
    \label{fig:unit_sphere}
\end{figure}

Parallel transport of a tangent vector along a closed loop $C$ in Riemannian geometry leads to its rotation with respect to the initial vector by the holonomy angle $\beta$ \cite{Arnold_1978}, commonly called the geometric phase in physics \cite{Unanyan_1999,Vitanov2017}.
Considering parallel transport along the surface of a unit (Bloch)) sphere in a three-dimensional Euclidean space, the geometric factor $\beta$ is equal to the solid angle $\Omega _C$ or the area of the surface subtended by the contour $C$. (Fig.~\ref{fig:unit_sphere}).
In this appendix, we demonstrate how the corresponding double integral  can be reduced to contour integral along the contour $C$.

First, let us project the spherical surface fragments  onto the $xy$-plane as shown in Fig.~\ref{fig:unit_sphere}.
In the spherical coordinates $\vector{R}=(R,\theta,\varphi)$ (with $R=1$ for the Bloch sphere), an infinitesimal element of the spherical surface is given by a solid angle $d{\Omega} = d{\theta}d{\varphi}\sin{\theta}$.
The projection $dS_{xy}=d{\Omega}\cos{\theta}$ of $d{\Omega}$ onto $xy$-plane allows us to rewrite $d{\Omega}$ as
\begin{eqnarray}
    \label{eq:ContourIntegeralEq1}
    d{\Omega} &= d{S_{xy}}/\cos{\theta} = d{r}d{\varphi}\frac{r}{\sqrt{1-r^{2}}}.
\end{eqnarray}
Here $r = \sin{\theta}$ is the projection of the radius $R=1$ onto $xy$-plane.

Let us assume the contour $C$ is sufficiently smooth, so that its $xy$ projection may be parametrized by the angle $\varphi$ as $C_{xy} = \{r\of{\varphi},\varphi\}$.
Then the geometric phase $\beta$ is obtained by integrating (\ref{eq:ContourIntegeralEq1}) over the area $S_{xy}$ bounded by $C_{xy}$:
\begin{eqnarray}
    \label{eq:ContourIntegeralEq2}
    \beta
    &=
    \int_{0}^{2\pi}{d\varphi}\int_{0}^{r\of{\varphi}}dr\frac{r}{\sqrt{1-r^{2}}}
    =
    2\pi - \int_{0}^{2\pi}d\varphi{\sqrt{1-r\of{\varphi}^{2}}}
\end{eqnarray}
While this relation formally is related to the contour $C_{xy}$, it can be transformed into a contour integral along $C$ by taking into account the following three properties:
(i) The angle $\varphi$ also paramatrizes the contour $C$ since the $z$-projection of the radius vector $\vector{R}\in{C}$ equals $z\of{\varphi} = \sqrt{1-r\of{\varphi}^{2}}$.
(ii) In spherical coordinates, the infinitesimal radius-vector increment can be expanded into the basis of coordinate  unit vectors $\vector{e}_\theta,\vector{e}_\varphi, \vector{e}_R$ as (Fig.~\ref{fig:unit_sphere})
\begin{eqnarray}
    \label{eq:ContourIntegeralEq3}
    d\vector{R} &= d\theta\vector{e}_{\theta} + d\varphi\sin{\theta}\vector{e}_{\varphi}+dR\vector{e}_{R} .
\end{eqnarray}
Therefore, $d\varphi = d\vector{R}\cdot\vector{e}_\varphi/\sin{\theta}$, which, (iii) taking into account the identity
\begin{eqnarray}
    \label{eq:ContourIntegeralProperty3}
    \vector{e}_\varphi = \vector{e}_z\times\vector{R}/\of{R\sin{\theta}},
\end{eqnarray}
allows us to rewrite (\ref{eq:ContourIntegeralEq2}) in the following form:
\begin{eqnarray}
    \label{eq:ContourIntegeralEq4}
    \beta
    &=
    2\pi -  I_C; \quad
  I_C=  \oint_{C}\left(d\vector{R}\cdot\vector{e}_z\times\vector{R}\right)\frac{z}{R\left(x^2+y^2\right)}
\end{eqnarray}
Here we have used the identity $\sin^{2}{\theta} = r^{2} = x^2+y^2$.

Contour integral $I_C$ in relation (\ref{eq:ContourIntegeralEq4}) corresponds to the expression (23) in work \cite{Unanyan_1999}, taking into account that
\begin{eqnarray}
    \label{eq:ContourIntegeralEq5}
    d\vector{R}\cdot\vector{e}_z\times\vector{R}
    &=
    \vector{e}_z\cdot \vector{R}\times d\vector{R}
    =
     xdy-ydx.
\end{eqnarray}

The relation (\ref{eq:ContourIntegeralEq4}) is not particularly convenient to use, since it contains a singularity along the whole z-axis $(-\infty < z < \infty)$ and the validity of its analytical continuation to contours not enclosing the $z$ axis may be questioned.
However, this singularity may be relaxed by confining it to the negative half-axis $(-\infty < z \leq 0)$ \cite{Mansuripur2017}.

From the relations (\ref{eq:ContourIntegeralEq3},\ref{eq:ContourIntegeralProperty3}), it follows that
\begin{eqnarray}
    \label{eq:ContourIntegeralEq6}
    \nabla\varphi
    &=
    \frac{\vector{e}_\varphi}{R\sin\theta}
    =
    \frac{\vector{e}_z\times\vector{R}}{R^{2}\sin^{2}\theta}.
\end{eqnarray}
Taking contour integral over $C$, we obtain a new identity:
\begin{eqnarray}
    \label{eq:ContourIntegeralEq7}
    2\pi
    &=
    \oint_{C}\left(d\vector{R}\cdot\vector{e}_z\times\vector{R}\right)\frac{1}{x^2+y^2}.
\end{eqnarray}
Inserting (\ref{eq:ContourIntegeralEq7}) into (\ref{eq:ContourIntegeralEq4}), and recognizing that $x^2+y^2=R^2-z^2$, we arrive at:
\begin{eqnarray}
    \label{eq:ContourIntegeralEq8}
    \beta
    &=
    \oint_{C}\left(d\vector{R}\cdot\vector{e}_z\times\vector{R}\right)\frac{R-z}{R\left(x^2+y^2\right)}
    =
    \oint_{C}\left(d\vector{R}\cdot\vector{e}_z\times\vector{R}\right)\frac{1}{R\left(R+z\right)}.
\end{eqnarray}
Clearly, the singularity in the integrand has been moved to the negative half-axis $(-\infty < z \leq 0)$.
The identity (\ref{eq:ContourIntegeralEq5}) may be inserted in the expression (\ref{eq:ContourIntegeralEq8}) to reproduce the relation (\ref{eq:E12}) after  identification of coordinates $x,y,z$ with coordinates $\Omega _S, \Omega _P, \Omega _Q$ associated with parameter space $\Re _3$ of lasers Rabi frequencies.

While the relation (\ref{eq:ContourIntegeralEq8}) was obtained from purely mathematical concepts, it has a deep physical meaning, related to the Dirac monopole \cite{Dirac1931,Mansuripur2017}.
The expression (\ref{eq:ContourIntegeralEq8}) has the form of a cyclic integral
\begin{eqnarray}
    \label{eq:ContourIntegeralEq9}
    \beta
    &=
    \oint_{C}d\vector{R}\cdot\vector{A};
    \quad
    \vector{A}
    =
    \frac{\vector{e}_z\times\vector{R}}{R\left(R + z\right)}
\end{eqnarray}
for the vector potential $\vector{A}$.
Dirac was the first to encounter similar $\vector{A}$ and a cyclic integral in his attempts \cite{Dirac1931} to describe the magnetic field $\vector{B}$ of a semi-infinite $(-\infty < z \leq 0)$ magnetized string. The straightforward calculations
\begin{eqnarray}
    \label{eq:ContourIntegeralEq10}
    \vector{B}
    &= \nabla\times\vector{A}
    =
    \vector{R}/R^3
\end{eqnarray}
show that the vector-potential corresponds to a magnetic monopole placed at the end of the semi-infinite string, while the Stokes theorem reveals that the geometrical factor $\beta$ equals the solid angle $\Omega$ of the spherical surface portion, subtended by the contour $C$.

\section{
    Fractional STIRAP infidelity $\bar{f}$ for harmonic laser pulses
    \label{appendix:InfidelityC}
}

The non-adiabatic coupling between bright and dark states leads to an outflow of population from the subspace of dark states, resulting in a drop in adiabatic passage efficiency. Here we estimate the corresponding infidelity parameter $\bar{f}$, assuming that under conditions of perfect adiabaticity, the influence of nonadiabatic processes does not strongly affect the temporal dynamics of the dark  state-vector $\ket{\Psi}$.

The matrix element $M_{DB}$ of the nonadiabatic coupling operator \cite{Shore2017,Wilczek1984} between  dark $\ket{\Psi}$ and bright $\ket{Br}$ states
\begin{eqnarray}
 \label{eq:MatrixElementC1}
 M_{DB}\equiv \hbar \bra{\widetilde{\Psi}_{\varphi }} \frac{\partial }{\partial t}\ket{Br _{\varphi }}=
 \im\hbar \frac{d \varphi}{d t} \bra{\widetilde{\Psi}_{\varphi }} \frac{\partial }{\partial \varphi }\ket{Br _{\varphi }}
\end{eqnarray}
has a simple geometrical representation when the parallel transport occurs along the circle $C_s$ depicted in Fig.~\ref{fig:ParallelTransportXYplane}. Since, as follows from the notation in  Fig.~\ref{fig:ParallelTransportXYplane},
\begin{eqnarray}
 \label{eq:RelationsC2}
 \frac{d \varphi}{d t}=\frac{2\pi }{T}; \quad \frac{\partial }{\partial \varphi }\ket{Br _{\varphi }}=r_s\vector{e}_{\varphi}; \quad (\widetilde{\Psi}_{\varphi } \vector{e}_{\varphi})=- \sin{\widetilde{\beta }_{\varphi}}
\end{eqnarray}
the matrix element (\ref{eq:MatrixElementC1}) reduces to
\begin{eqnarray}
 \label{eq:MatrixElementC3}
 M_{DB}= -\im\hbar r_s \frac{2\pi }{T}\sin{\widetilde{\beta }_{\varphi}}
\end{eqnarray}
where  $\widetilde{\beta }_{\varphi}$ is the current rotation angle (\ref{eq:geometricPhaseBetaG}).

Noteworthy, the adiabacity implies that a subspace of dark states in Fig.~\ref{fig:tripod}b can be replaced by the single state-vector $\ket{\Psi}$. A corresponding linkage diagram is identified thus with a simple $\Lambda$-scheme with $\ket{Br}$ as an intermediate state coupled to $\ket{2}$ and $\ket{\Psi}$ levels by $\Omega _{eff}$ and $2M_{DB}/\hbar$ Rabi frequencies consequently. The adiabaticity criterion (\ref{eq:E11}) allows us to apply for that $\Lambda$-scheme the adiabatic elimination procedure resulting in the  formation of the population-trapped state \cite{Shore2017} with the following probability amplitudes: $C_{Br}\cong  0$
\begin{eqnarray}
 \label{eq:TrappedStateC4}
 C_{2}\cong \frac{2M_{DB}}{\hbar \Omega _{eff}}=-\im\frac{2\pi}{T\Omega _{eff}/2}r_s\sin{\widetilde{\beta}_\varphi  };
\quad
 C_{\Psi}\cong  \sqrt{1-|C_2|^2}.
\end{eqnarray}

The infidelity parameter (\ref{eq:InfidelityE38}) $\bar{f}=1-C_{\Psi}^2=|C_2|^2$  is identical in our case to the population of state 2 at the final moment of adiabatic passage, i.e., when $\varphi=2\pi$. To find an explicit formula for $\bar{f}$, consider the following three points: (i) the  adiabacity parameter  $\wp =T\Omega _{eff}/2$ (see Eq.~(\ref{eq:ParametersE34})); (ii) relation (17)
gives $\sin{\widetilde{\beta}_{2\pi}}=\sin{\beta}$; (iii) expressions (\ref{eq:betaRelationTolambda}) results in $(2\pi r_s)^2=(4\pi -\beta)\beta$. Combining equation~(\ref{eq:TrappedStateC4}) with the above points (i)-(iii), we get the expression (\ref{eq:AmplitudeC_2}) for the $C_2$ amplitude
along with the analytical result~(\ref{eq:InfidelityE40}) for STIRAP infidelity.

\section*{References}
\bibliographystyle{iopart-num}
\bibliography{adiabatic-passage-in-tripod}

\providecommand{\noopsort}[1]{}\providecommand{\singleletter}[1]{#1}%
\providecommand{\newblock}{}
\begin{thebibliography}{10}
\expandafter\ifx\csname url\endcsname\relax
  \def\url#1{{\tt #1}}\fi
\expandafter\ifx\csname urlprefix\endcsname\relax\def\urlprefix{URL }\fi
\providecommand{\eprint}[2][]{\url{#2}}

\bibitem{Bergmann_2019}
Bergmann K, Nägerl H~C, Panda C, Gabrielse G, Miloglyadov E, Quack M, Seyfang
  G, Wichmann G, Ospelkaus S, Kuhn A, Longhi S, Szameit A, Pirro P, Hillebrands
  B, Zhu X~F, Zhu J, Drewsen M, Hensinger W~K, Weidt S, Halfmann T, Wang H~L,
  Paraoanu G~S, Vitanov N~V, Mompart J, Busch T, Barnum T~J, Grimes D~D, Field
  R~W, Raizen M~G, Narevicius E, Auzinsh M, Budker D, P{\'{a}}lffy A and Keitel
  C~H 2019 {\em Journal of Physics B: Atomic, Molecular and Optical Physics\/}
  {\bf 52} 202001 \urlprefix\url{https://doi.org/10.1088/1361-6455/ab3995}

\bibitem{Shore2017}
Shore B~W 2017 {\em Adv. Opt. Photon.\/} {\bf 9} 563--719
  \urlprefix\url{http://opg.optica.org/aop/abstract.cfm?URI=aop-9-3-563}

\bibitem{Unanyan_1999}
Unanyan R~G, Shore B~W and Bergmann K 1999 {\em Phys. Rev. A\/} {\bf 59}(4)
  2910--2919 \urlprefix\url{https://link.aps.org/doi/10.1103/PhysRevA.59.2910}

\bibitem{Vitanov2017}
Vitanov N~V, Rangelov A~A, Shore B~W and Bergmann K 2017 {\em Rev. Mod.
  Phys.\/} {\bf 89}(1) 015006
  \urlprefix\url{https://link.aps.org/doi/10.1103/RevModPhys.89.015006}

\bibitem{Vitanov2013}
Rousseaux B, Gu\'erin S and Vitanov N~V 2013 {\em Phys. Rev. A\/} {\bf 87}(3)
  032328 \urlprefix\url{https://link.aps.org/doi/10.1103/PhysRevA.87.032328}

\bibitem{Berry_1984}
Berry M~V 1984 {\em Proceedings of the Royal Society of London. A. Mathematical
  and Physical Sciences\/} {\bf 392} 45--57 (\textit{Preprint}
  \eprint{https://royalsocietypublishing.org/doi/pdf/10.1098/rspa.1984.0023})
  \urlprefix\url{https://royalsocietypublishing.org/doi/abs/10.1098/rspa.1984.0023}

\bibitem{Wilczek1984}
Wilczek F and Zee A 1984 {\em Phys. Rev. Lett.\/} {\bf 52}(24) 2111--2114
  \urlprefix\url{https://link.aps.org/doi/10.1103/PhysRevLett.52.2111}

\bibitem{Kreyszig1991}
Kreyszig E 1991 {\em Differential Geometry\/} Differential Geometry (Dover
  Publications) ISBN 9780486667218
  \urlprefix\url{https://books.google.lv/books?id=P73DrhE9F0QC}

\bibitem{Arnold_1978}
Arnold V~I 1978 {\em Mathematical Methods of Classical Mechanics\/} (New York,
  NY: Springer New York) ISBN 978-1-4757-1693-1

\bibitem{Fadeev_1980}
Faddeev L~D and Slavnov A~A 1980 {\em Gauge Fields, Introduction to Quantum
  Theory\/} (The Benjamin/Cummings Publishing Company, Inc.) ISBN 0-8053-9016-2

\bibitem{Dykhne1960}
Dykhne A 1960 {\em Sov. Phys. JETP\/} {\bf 11} 411

\bibitem{DavisPechukas1976}
Davis J~P and Pechukas P 1976 {\em The Journal of Chemical Physics\/} {\bf 64}
  3129--3137 (\textit{Preprint}
  \eprint{https://aip.scitation.org/doi/pdf/10.1063/1.432648})
  \urlprefix\url{https://aip.scitation.org/doi/abs/10.1063/1.432648}

\bibitem{Dirac1931}
Dirac P~A~M 1931 {\em Proceedings of the Royal Society of London. Series A,
  Containing Papers of a Mathematical and Physical Character\/} {\bf 133}
  60--72 (\textit{Preprint}
  \eprint{https://royalsocietypublishing.org/doi/pdf/10.1098/rspa.1931.0130})
  \urlprefix\url{https://royalsocietypublishing.org/doi/abs/10.1098/rspa.1931.0130}

\bibitem{Unanyan1998}
Unanyan R, Fleischhauer M, Shore B and Bergmann K 1998 {\em Optics
  Communications\/} {\bf 155} 144--154 ISSN 0030-4018
  \urlprefix\url{https://www.sciencedirect.com/science/article/pii/S0030401898003587}

\bibitem{Kirova2017}
Kirova T, Cinins A, Efimov D~K, Bruvelis M, Miculis K, Bezuglov N~N, Auzinsh M,
  Ryabtsev I~I and Ekers A 2017 {\em Phys. Rev. A\/} {\bf 96}(4) 043421
  \urlprefix\url{https://link.aps.org/doi/10.1103/PhysRevA.96.043421}

\bibitem{Landau_QM1981}
Landau L and Lifshitz E 1981 {\em Quantum Mechanics: Non-Relativistic Theory\/}
  Course of Theoretical Physics (Butterworth-Heinemann) ISBN 9780080503486
  \urlprefix\url{https://books.google.lv/books?id=SvdoN3k8EysC}

\bibitem{Aharonov1987}
Aharonov Y and Anandan J 1987 {\em Phys. Rev. Lett.\/} {\bf 58}(16) 1593--1596
  \urlprefix\url{https://link.aps.org/doi/10.1103/PhysRevLett.58.1593}

\bibitem{Mansuripur2017}
Mansuripur M 2017  \urlprefix\url{https://arxiv.org/abs/1701.00592}

\end{thebibliography}

\end{document}